\documentclass[prd,aps,floats,preprintnumbers,preprint,nofootinbib,groupedaddress,11pt]{revtex4}
\def\eslt{\not\!\!{E_T}} 
\def\eslt{E_T^{\rm miss}}
\def\ba{\begin{array}} 
\def\ea{\end{array}}
\def\to{\rightarrow}

\def\bi{\begin{itemize}}
\def\ei{\end{itemize}} 
 
\def\th{\tilde h}

\def\tq{\tilde q} 
 
\def\tz{\widetilde Z}

 \def\alt{\stackrel{<}{\sim}}
 
\def\be{\begin{equation}}
\def\ee{\end{equation}} 
\def\shat{\hat{S}}

\usepackage{amsmath}
\usepackage{graphicx}

\begin{document}

\preprint{UH-511-1139-09}

\title{ Addressing  $\mu-b_\mu$ and proton lifetime problems and
  active neutrino masses in a $U(1)'$-extended supergravity model}

\author{R.~S. Hundi}
 \email{srikanth@phys.hawaii.edu}
\author{Sandip Pakvasa}%
\email{pakvasa@phys.hawaii.edu}
\author{Xerxes Tata}%
 \email{tata@phys.hawaii.edu}
\affiliation{%
\mbox{Dept. of Physics and Astronomy}, University of Hawaii, Honolulu, HI 96822, U.S.A.\\
}%

\begin{abstract}
\noindent We present a locally supersymmetric extension of the minimal
supersymmetric Standard Model (MSSM) based on the gauge group
$SU(3)_C\times SU(2)_L\times U(1)_Y\times U(1)'$ where, except for the
supersymmetry breaking scale which is fixed to be $\sim 10^{11}$~GeV, we
require that all non-Standard-Model parameters allowed by the {\it
local} spacetime and gauge symmetries assume their natural values. The
$U(1)'$ symmetry, which is spontaneously broken at the intermediate
scale, serves to ({\it i})~explain the weak scale magnitudes of $\mu$
and $b_\mu$ terms, ({\it ii})~ensure that dimension-3 and dimension-4
baryon-number-violating superpotential operators (and, in a class of
models, {\it all} $\Delta B=1$ operators) are forbidden, solving
the proton-lifetime problem, ({\it iii}) predict {\it bilinear lepton
number violation} in the superpotential at just the right level to
accommodate the observed mass and mixing pattern of active neutrinos
(leading to a novel connection between the SUSY breaking scale and
neutrino masses), while corresponding trilinear operators are strongly
supppressed. The phenomenology is like that of the MSSM with bilinear
$R$-parity violation, were the would-be lightest supersymmetric particle
decays leptonically with a lifetime of $\sim
10^{-12}-10^{-8}$~s. Theoretical consistency of our model requires the
existence of multi-TeV, stable, colour-triplet, weak-isosinglet scalars
or fermions, with either conventional or exotic electric charge which
should be readily detectable if they are within the kinematic reach of a
hadron collider. Null results of searches for heavy exotic isotopes
implies that the re-heating temperature of our Universe must have
been below their mass scale which, in turn, suggests that sphalerons
play a key role for baryogensis. Finally, the dark matter cannot be the
weakly interacting neutralino.
\end{abstract}

\maketitle

\section{Introduction} \label{intro}

Softly broken supersymmetry (SUSY) with weak scale super-partners is a
theoretically appealing and phenomenologically viable framework for
physics beyond the Standard Model (SM) \cite{susy-text}. Weak scale SUSY
provides an elegant mechanism to stabilize the weak interaction scale
against runaway quantum corrections that arise when the SM is embedded
into a larger framework that includes particles much heavier than the
weak scale \cite{hier}.  As a result, SUSY models provide a much more
convincing setting for the unification of the strong and electroweak
interactions of the SM into a single interaction at the much larger
scale $M_{\rm GUT}$ that appears in grand unified theories
(GUTs). Moreover, it is well-known that the measured value of gauge
couplings (and of the down-type third generation fermion masses) are
incompatible with grand unification if these are extrapolated to high
energy as in the SM, but unify rather well in SUSY GUTs with
super-partners of SM particles around the TeV scale \cite{unifi}.  Also,
SUSY theories with a conserved $R$-parity quantum number can readily
accommodate the observed amount of cold dark matter, most naturally
(though not necessarily) in the form of a weak interacting massive
neutralinos that are left over as {\it thermal relics} from the Big Bang
\cite{susy-cos}.

While these remarkable properties of SUSY have continued to provide
impetus for its exploration even in the absence of any direct evidence
from searches at high energy colliders -- a situation that, we hope,
will change once the data from the Large Hadron Collider (LHC) become
available -- we note that {\it generic} \ SUSY models give rise to new
problems not present in the SM. These include:
\begin{itemize}
\item The $\mu$-problem: Why is the coefficient of the gauge-invariant
  ${\hat{H}}_u{\hat{H}}_d$ superpotential term not as large as the GUT
  scale but of about the weak scale as needed for phenomenology? In
  addition, we also need a soft SUSY breaking (SSB) scalar bilinear term,
  with its coefficient $b_\mu$ also taking on a weak scale value.
\item The proton decay problem: Why are the renormalizable $SU(3)\times
  SU(2)\times U(1)_Y$ invariant baryon- and lepton-number violating
  couplings that could potentially cause weak scale proton decay small
  (or, more likely, absent)? In addition, why are the couplings of
  $R$-parity conserving dimension-4 baryon and lepton number violating
  operators in the superpotential also small enough so as not to conflict
  with the limts on proton lifetime?
\item The SUSY flavour and $CP$ problems: Why are quark/lepton
  flavour-violating and $CP$-violating couplings so much smaller than
  their naturally expected values?
\end{itemize}
We stress that these are issues only for a generic SUSY model in that a
number of mechanisms to evade each of these ``problems'' have been
suggested in the literature \cite{mupro,prodecay,flav}.

We speculate that the answers to these
questions will be evident once the mechanisms by which the dimensionful
parameters in the sparticle sector arise are understood. Our goal here
is to present a new model that addresses the first two of the three
problems mentioned above (we have nothing new to add about flavour or
$CP$ violation), where gravitational interactions convey the effects of
SUSY-breaking that occurs in a ``hidden sector'' to the ``observable
sector'' which includes the SM particles and their superpartners,
together with additional exotics (some of which may be close to the weak
scale) that we are forced to include for the
consistency of the framework. 

In view of the fact that there are already numerous
models that attempt to address one or more of the issues, we should
explain our rationale for constructing yet one more model. The main
reason is that for our construction we adopt the following reasonable
ground rules that are not all respected by other
authors. 
\begin{enumerate}
\item We present the complete dynamics of both SUSY breaking and
  its mediation to the observable sector that determines the various
  scales in the theory. In other words, we do not
  simply assume that certain fields get vacuum expectation values
  ($vevs$) at appropriate scales. 

\item Since there are arguments \cite{globalgrav} that suggest that
  gravitational dynamics does not respect global symmetries, we allow
  ourselves to use only gauge symmetries to restrict the form of the
  dynamics; {\it i.e.} we eschew {\it ad hoc} global symmetries,
  including any $R$-symmetry.  In other words, the weak scale values for
  $\mu$ and $b_\mu$, as well as the observed lower bound on the
  life-time of the proton are derived as a consequence of {\it local}
  symmetries.

\item All (non-SM) interactions not explicitly forbidden by the
  symmetries are assumed to be allowed, and with one exception discussed
  below, with natural values for the parameters. We have, of course, no
  explanation of the pattern of SM Yukawa couplings which, as usual, we
  take to reproduce the observed fermion masses.
\end{enumerate}

It is well known that in all realistic models where gravity acts as the
messenger of SUSY-breaking to the observable sector, SUSY breaking
occurs at an intermediate scale $\Lambda$ such that $\Lambda^2/M_{\rm P}
\sim 1$~TeV, where $M_{\rm P}$ is the Planck scale. The small ratio,
$\Lambda/M_{\rm P}$ is usually unexplained.\footnote{There are, however,
suggestions where this hierarchy may be accounted for by
non-perturbative dynamics.} Our attempt is not different in this respect
in that we also set the scale of SUSY-breaking by hand to be
hierarchically different from the Planck scale. The novel feature of our
model is that this same intermediate scale $\Lambda$ sets the scale of
active neutrino masses (along with the scale of the $\mu$ and $b_\mu$
parameters), and that it is possible to accommodate -- but not explain
-- the observed mixing pattern of neutrinos.

The remainder of this paper is organised as follows.  In
Sec.~\ref{sec:prel} we present the general ideas behind how we obtain
the SSB parameters, weak scale $\mu$ and $b_\mu$ terms and neutrino
masses. The construction of the model is completed in
Sec.~\ref{sec:complete} where several exotic fields necessary to cancel
quantum anomalies are introduced. The broad aspects of the phenomenology of the
model are discussed in Sec.~\ref{sec:phen}. These include, the
suppression of proton decay and $n\bar{n}$ oscillations, neutrino masses
and mixing, the spectrum of new particles and their signals at the LHC,
and finally some cosmological considerations. We summarize our findings
in Sec.~\ref{sec:summ}.

\section{Model Preliminaries}\label{sec:prel}
We begin the construction of our supergravity-based framework, focussing
for the moment only on general features -- the new fields, and the
origin of associated scales that are essential for viable
phenomenology. Discussion of some details necessary in order to obtain
the complete model is deferred to Section~\ref{sec:complete}. 

The general approach for solutions to the $\mu$ problem is to include a
new symmetry, perhaps an $R$ symmetry, that forbids the introduction of
the superpotential $\mu$-term which is then generated only upon the
spontaneous breakdown of this symmetry \cite{mupro}. We take the same
approach and, for reasons explained above, introduce a new $U(1)'$ gauge
symmetry that forbids the $\mu$ term. We will arange the dynamics so
that spontaneous breakdown of this local symmetry generates both the
$\mu$ as well as the SSB $b_\mu$ terms with weak scale values. We begin,
however, with the supersymmetry breaking sector and the generation of
SSB terms for the superpartners of the SM particles that results from
the gravitational coupling between the supersymmetry-breaking sector and
the SM superfields.

\subsection{Supersymmetry breaking}

As usual, we assume that supersymmetry is broken in a hidden sector that
couples to SM particles and their superpartners only via (very
suppressed) gravitational interactions \cite{sugra}. Following Polonyi
\cite{polonyi}, we introduce a superfield $\shat$ which is a singlet
under both the SM gauge group as well as the new $U(1)'$ symmetry that,
as we said, precludes us from including the $\mu$-term in the
superpotential. Since there are no symmetry considerations to restrict
the self couplings of $\shat$ in the superpotential, we must allow the
hidden sector potential to be an arbitrary function of $\shat$, and not
restrict it to be the linear Polonyi superpotential. Since we are
talking about the effective theory at the Planck scale, we would expect
that $M_P$ determines the scale of the superpotential for $\shat$.
However, in order to obtain the SUSY breaking $vev$ $\sqrt{{\cal F}_S}$
at the intermediate scale $\Lambda$, and to ensure the subsequent 
cancellation of the cosmological constant, we are forced to
choose the overall scale of the superpotential to be much smaller than
$M_P$ (this {\it ad hoc} choice of scale, is the exception
mentioned in item 3. of Sec.~\ref{intro}, and is common to most
supergravity models), so that we write
\begin{equation}
{\hat{W}}_{0}(\hat{S}) = \frac{\Lambda^2}{M_P^2}\left[M_P^2\hat{S}+\alpha M_P
\hat{S}^2+\beta \hat{S}^3
+\gamma M_P^3\right],
\label{eq:hid}
\end{equation}
with the dimensionless coefficients $\alpha$, $\beta$ and $\gamma$ being
${\cal O}(1)$, and $\Lambda$ determining the over-all scale of this
superpotential.  The precise details of the form of ${\hat{W}}_0$ are
unimportant as long as the mass parameters and $vev$s (if non-zero) are
${\cal O}({M_P})$. We terminate the series after the cubic term
only for definiteness. Naive dimensional analysis suggests that if it
does not vanish, $\langle S \rangle \sim M_P$, while ${\cal F}_S \sim
\Lambda^2$. 

\subsection{Soft symmetry breaking terms for MSSM superpartners}

The MSSM \cite{dgsak} includes the
superfields,
\begin{equation}
{\hat{\Phi}}_i =
(\hat{Q}_\bullet,\hat{D}^c_\bullet,\hat{U}^c_\bullet,\hat{L}_\bullet,
\hat{E}^c_\bullet, \hat{H}_d,\hat{H}_u),
\label{eq:supphi}
\end{equation} where the $\bullet$ denotes
the generation index. These fields constitute the MSSM sector. We write
down the most general superpotential consistent with the {\it local}
$SU(3)_C\times SU(2)_L\times U(1)_Y\times U(1)'$ gauge symmetry, where,
as mentioned above, the last factor is the new local $U(1)$ symmetry
that we introduce to forbid the $\hat{H}_u\hat{H}_d$ term in the
superpotential.  We will see below that we can assign $U(1)'$ charges
consistently with the cancellation of gauge and gravitational anomalies,
so that {\it the same $U(1)'$ symmetry also forbids dimension-2,
dimension-3, dimension-4 baryon-number and lepton-number violating
operators involving the MSSM fields in the superpotential.} The
renormalizable MSSM superpotential, therefore, includes only the usual
fermion Yukawa coupling terms. The $\mu$-term, together with other
dimensionful terms (discussed below) will be dynamically generated.

We choose the superpotential, the K\"ahler potential and the gauge
kinetic function for our effective theory valid below the Planck scale
to be,
\begin{eqnarray}
\hat{W} &=& \hat{W}_{0}(\hat{S}) + \hat{W}_{\rm MSSM} + \cdots,
	\nonumber \\ \hat{K} &=& \hat{S}^\dagger \hat{S} +
	\sum_i\hat{\Phi}^\dagger_i\hat{\Phi}_i + \sum_i
	\left(K_i\left(\frac{\hat{S}}{M_P},\frac{\hat{S}^\dagger}{M_P}\right)\hat{\Phi}^\dagger_i
	\hat{\Phi}_i+ h.c.\right) + \cdots, \label{eq:WKf}\\
	\hat{f}_{(\alpha)AB} &=&
	\delta_{AB}\left(1+f_{(\alpha)}\frac{\hat{S}}{M_P} +
	\cdots\right) \nonumber,
\end{eqnarray}
where the ellipses denote terms involving other fields which we will
introduce later, or Planck-suppressed higher order terms that are
undoubtedly present but will not alter the SSB masses and couplings of
the MSSM fields.  Here, $f_{(\alpha)}$ are dimensionless parameters
taken to be ${\cal O}$(1), $A,B$ label the gauge group indices while the
index $\alpha$ labels the gauge group factor [$SU(3)_C$,
$SU(2)_L$,$U(1)_Y$ or $U(1)'$], and $\hat{W}_{\rm MSSM}$ specifies the
usual superpotential Yukawa couplings of the MSSM superfields. The last
term involving the functions $K_i$ in the K\"ahler potential generically
results in non-universal SSB mass parameters for the MSSM fields when
the scalar component of $\hat{S}$ acquires a $vev \sim M_P$
\cite{soniweld}.

The scalar potential in supergravity is
\begin{equation}
V = M_P^4e^G(G_i(G^{-1})^i_jG^j -3)+\frac{1}{2}M_P^4g^2\left(
{\rm Re}f^{-1}_{AB}\right)G^i(t_A)_{ij}\phi_jG^k(t_B)_{kl}\phi_l,
\label{eq:sugsca}
\end{equation}
where, following the notation of Ref.\cite{wss}, 
\begin{eqnarray}
G &=& \left(\frac{\hat{K}}{M_P^2} +
	\ln|\frac{\hat{W}}{M_P^3}|^2\right)_{\hat{\Phi}_i=\phi_i}, \nonumber
	\\ 
G^i &=& \left.\frac{\partial
	G}{\partial\hat{\Phi}_i}\right|_{\hat{\Phi}_i=\phi_i}, \quad G_j
	= \left.\frac{\partial
	G}{\partial\hat{\Phi}^{j\dagger}}\right|_{\hat{\Phi}_i=\phi_i},
\label{eq:G}	 \\ 
G^i_j &=& \left.\frac{\partial^2G}{\partial\hat{\Phi}_i
	\partial\hat{\Phi}^{j\dagger}}\right|_{\hat{\Phi}_i=\phi_i}.\nonumber
\end{eqnarray}
In this expression for the scalar potential of a general locally
supersymmetric quantum field theory, we have abused notation and used
$\Phi_i$ to denote {\it any superfield}, and $\phi_i$ the scalar
component of $\hat{\Phi}_i$. We trust that the dual use of $\Phi_i$ as
any superfield here, and also as the symbol for the chiral superfields
of the MSSM in (\ref{eq:supphi}), will not cause any confusion. In the
last term of (\ref{eq:sugsca}), $g$ denotes the gauge coupling constant,
in general different for each of the four gauge group factors, and $t_A$ denote the
generators of the gauge group. Note that in this term in the scalar
potential, we have suppressed the index $\alpha$ (on $g$, on the gauge
kinetic function, and on the gauge group generators) which is implicitly
summed over all four gauge group factors.  Substituting (\ref{eq:WKf})
into the equation for the scalar potential gives us the SSB parameters
for the MSSM fields. Because of the non-minimal terms involving $K_i$ in
the K\"ahler potential, we must rescale the fields (by non-universal
factors $\sim 1$) so that the kinetic energy terms take their canonical
form in order to read of the scalar SSB mass parameters and trilinear
couplings.  The scale for these soft terms, which would have taken on a
universal value had we not required the rescaling of fields just
mentioned, can be written as,
\begin{equation}
m_{\rm soft} \sim \frac{\langle \hat{W}|_{\hat{\Phi}_i=\phi_i}\rangle}{M_P^2}
\simeq \frac{\langle \hat{W}_0|_{\hat{S}=S}\rangle}{M_P^2}.
\label{eq:sof}
\end{equation}
Since $\langle S\rangle\sim M_P$, we have $\langle
\hat{W}_{0}\rangle\sim\Lambda^2M_P$, so that we must choose $\Lambda\sim
10^{10}-10^{11}$~GeV in order for $m_{\rm soft}$ to be at the TeV
scale. This well known reasoning applies to matter and Higgs scalar mass
parameters as well as to the trilinear interactions, but not to the SSB
$b_\mu$ term which, like the supersymmetric $\mu$-term, is forbidden by
the $U(1)'$ symmetry.

Gaugino masses arise because the gauge kinetic functions are
field-dependent \cite{susy-text}. The gaugino mass matrix
(which is, of course, diagonal) is generically given by,
\begin{equation}
m_\lambda=\frac{1}{2}M_Pe^{G/2}\left.\frac{\partial \hat{f}^*_{AB}}{\partial\Phi^{j*}}\right|_{\Phi^{j*}=\phi^{j*}}
(G^{-1})^j_kG^k.
\end{equation}
Since $G^k$ and $\left.\frac{\partial
  \hat{f}^*_{AB}}{\partial\Phi^{j*}}\right|_{\Phi^{j*}=\phi^{j*}}\sim
  1/M_P$, $(G)^j_k\sim M_P^2$ with $e^{G/2} \sim \Lambda^2/M_P^2$, the
  magnitude of the gaugino mass parameters is $\sim \Lambda^2/M_P$ which
  is comparable to the other SSB parameters as desired. If the gauge
  kinetic function depends on the gauge group factor (through, {\it
  e.g.} $f_{(\alpha)}$ in (\ref{eq:WKf})), non-universal gaugino masses
  will result.

This discussion of SSB parameters in supergravity models is not new. We present
it mainly to set up notation, and for the sake of completeness.

\subsection{$\mu$ and $b_\mu$ terms} \label{subsec:mu}

To explain why the $\mu$-term has a magnitude around the weak scale
rather than $M_P$, we choose $U(1)'$ charges so that the operator
$\hat{H}_u\hat{H}_d$ is forbidden in the superpotential. An effective
$\mu$-term is then generated either via the $vev$ of the auxiliary
component of a new (elementary or composite) SM singlet superfield
$\hat{Z}^\dagger$ that spontaneously breaks $U(1)'$ and couples to the
MSSM Higgs superfields in the K\"ahler potential \cite{GMM}, or via
the $vev$ of the scalar components of $\hat{Z}$ with a superpotential
coupling to $\hat{H}_u\hat{H}_d$ \cite{KNM}. In either case, we
have to ensure that an SSB breaking $b_\mu$ term, also with a weak
scale magnitude, can be generated consistent with the assumed local
symmetry.

Guidice and Masiero \cite{GMM} proposed that $\mu$ may be generated via
the term
$$\hat{K}\ni \frac{\hat{Z}^{\dagger n}\hat{H}_u\hat{H}_d}{M_P^n}$$ which
would lead to $\mu \sim {{\cal F}_Z Z^{n-1}}/{M_P^n}$,
where we have abused notation to denote the $vev$s of the auxiliary and
scalar components of $\hat{Z}$ by the corresponding fields. To generate
a non-zero value for ${\cal F}_Z$, we must have \footnote{The $m$
$\hat{Z}'$ fields in this superpotential need not all be the same. In
other words, for $m=2$ the superpotential operator could be
$\hat{Z}\hat{Z}^{'}_1\hat{Z}^{'}_2$. To obtain the following estimates
assume that the $vev$s of the three fields (if non-zero) have comparable
magnitudes.}, $$\hat{W}\ni \frac{\hat{Z} \hat{Z}^{'m}}{M_P^{m-2}},$$
which suggests ${\cal F}_Z \sim \frac{Z^{'m}}{M_P^{m-2}} \sim
\frac{Z^{m}}{M_P^{m-2}}$. These required couplings of $\hat{Z}$ and
$\hat{Z}'$ must, of course be consistent with their $U(1)'$ charges. In
this case, it is easy to see that we must also include
$$\hat{W}\ni \frac{\hat{Z}^{'mn}\hat{H}_u\hat{H}_d}{M_P^{mn-1}}$$ in the
superpotential since it is not forbidden by the $U(1)'$ symmetry. A
$vev$ for $Z^{'}$ will amend the magnitude of $\mu$ from its value above
by a factor $1+\left(\frac{Z}{M_P}\right)^{(m-1)(n-1)} \sim {\cal O}(1)$ for $m,n\ge
1$. This new contribution can potentially also give a contribution to
$b_\mu$ which is $\sim
\left(\frac{Z}{M_P}\right)^{(m-1)(n-1)-n}\times\mu^2$. We see that the
choice $n=1$ in the original Guidice-Masiero proposal potentially gives
an undesirably large value  $b_\mu \sim \mu^2 \times \frac{M_P}{Z}$.
Only values of $m$ and $n$ such that $(m-1)(n-1) \ge n$ are guaranteed
to be ``safe'' in this regard. 

Alternatively, we can generate $\mu$ via a superpotential term,
$$\hat{W}\ni \frac{\hat{Z}^n\hat{H}_u\hat{H}_d}{M_P^{n-1}}$$ which will
give $\mu\sim \frac{Z^n}{M_P^{n-1}}$ if the scalar component of $Z$
acquires a $vev$ \cite{KNM}. If ${\cal F}_Z$ also acquires a
$vev$ via a superpotential term,
$$\hat{W} \ni \frac{\hat{Z}^p\hat{Z}^{'q}}{M_P^{p+q-3}},$$ we obtain
$b_\mu \sim \mu^2 \times \left(\frac{Z}{M_P}\right)^{p+q-n-2} \sim
\mu^2$ if $p+q=n+2$.  If $\frac{qn}{p}$ is not an integer, the $U(1)'$
symmetry precludes the appearance of the
operator $\hat{Z}^{'\dagger P}\hat{H}_u\hat{H}_d$ in the K\"ahler
potential for any integer value of $P$, so that there can be no corresponding
contribution to $\mu$ via the Guidice-Masiero mechanism.

In the following, we will use this second mechanism with $n=2, \  p=3$ and
$q=1$ to generate weak scale values for both $\mu$ and $b_\mu$. This
then requires that $\langle Z\rangle \sim \langle Z'\rangle \sim \Lambda
\sim 10^{11}$~GeV.  Moreover, we will see that the $vev$ for the scalar
component of the superfield $\hat{Z}^{'}$ that we are led to introduce
to get a non-zero auxiliary component of $\hat{Z}$ also serves to give
the desired mass scale in the neutrino sector.

\subsection{Neutrino masses}
\label{sec:C}
It is well known \cite{LV-Rp} that lepton number and $R$-parity violating terms in
the superpotential 
\begin{equation}
\hat{W}_{LV} \ni \lambda_{ijk} \epsilon_{ab}\hat{L}_i^a\hat{L}_j^b\hat{E}^c_k + \lambda^\prime_{ijk}\epsilon_{ab}
\hat{L}_i^a\hat{Q}_j^b\hat{D}^c_k + \epsilon_i
\epsilon_{ab}\hat{L}_i^a\hat{H}_u^b 
\label{eq:LV}
\end{equation}
induce masses for active neutrinos which then are Majorana
fermions. Here, $a$ and $b$ are $SU(2)_L$ indices, and we have
suppressed colour indices on $\hat{Q}$ and $\hat{D}^c$ in the second
term.  The dimensionless coupling constants $\lambda_{ijk}$
($\lambda_{ijk}^\prime$) are antisymmetric in the generation indices $i$
and $j$ ($j$ and $k$).  The last term in (\ref{eq:LV}) evidently leads
to mixing between the active neutrino fields and the higgsinos,
resulting in a $7\times 7$ neutrino--higgsino--neutral-gaugino mass
matrix. Assuming that $\epsilon_i$ are all much smaller than the other
entries of this matrix (which all have a weak scale magnitude) of this
matrix, we see that one linear combination of the neutrinos acquires a
mass $\sim {\epsilon_i^2\over M_{\rm weak}}$ at the {\it tree-level},
while other neutrinos acquire masses via radiative corrections since
there is no symmetry that precludes this.  For $|\epsilon_i| \sim
10^{-4}$~GeV, we see that the tree-level neutrino mass scale is $\sim
0.1$~eV. 

Just as for the $\mu$-parameter, we have to explain  why the magnitude of
$\epsilon_i$ is so small. We envisage that this bilinear term is
forbidden in the superpotential, and arises only when
the scalar component of a superfield $\hat{X}$ that enters the
superpotential via the dimension-5 operator (with smaller powers of
$\hat{X}$ being forbidden by the $U(1)'$ symmetry), 
 \begin{equation}
\hat{W}\ni \frac{\hat{X}^3}{M_P^2}\hat{L}\hat{H}_u\;,
\label{eq:bi}
\end{equation}
acquires a $vev$, spontaneously breaking 
the $U(1)'$ symmetry that we have already
introduced to alleviate the $\mu$ problem. Remarkably, we see that if
$\hat{X}\sim 10^{11}$~GeV, the desired magnitude for $\epsilon_i$ is
obtained. 

\begin{figure}
\begin{center}
\includegraphics[]{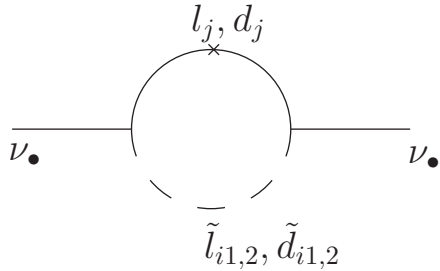}
\end{center}
\caption{Diagrams by which neutrino masses are generated at the one loop
level by the trilinear $R$-parity violating superpotential interactions
in (\ref{eq:LV}).}
\label{fig:loop}
\end{figure}

As we have mentioned, neutrino mass matrices may also be generated 
via the operators
$\hat{L}_i\hat{L}_j\hat{E}^c_k$ and $ \hat{L}_i\hat{Q}_j\hat{D}^c_k$ which 
violate lepton number by one
unit. They generate a neutrino mass matrix at one loop level via the diagrams
shown in Fig.~\ref{fig:loop}.
Loops with third generation leptons/quarks yield the largest entries,
and the corresponding scale of the neutrino mass is given by,
\begin{equation}
m_\nu\sim\frac{\lambda\cdot\lambda}{32\pi^2}\sin 2\theta_\tau m_\tau
 \ \ {\rm or} \ \ \frac{\lambda^{\prime }\cdot\lambda^{\prime}}{32\pi^2}\sin 2\theta_b m_b,
\end{equation}
where $\lambda$ and $\lambda'$ denote the appropriate $\lambda_{ijk}$ or
$\lambda'_{ijk}$ coupling, and $\theta_\tau ,\theta_b$ are the
intragenerational mixing angles for tau sleptons and bottom squarks,
respectively.  Since $\sin 2\theta_f \sim {m_f\over M_{\rm SUSY}} \sim
10^{-2}$ for $f=b,\tau$, we see that we can obtain a neutrino mass scale
of 0.1~eV (which is consistent with all data) if $\lambda ,
\lambda^\prime\sim 10^{-3}$.

In keeping with our stated philosophy, a coupling of this magnitude
requires explanation. We may envision that these couplings, which are
forbidden at the tree-level by the $U(1)'$ symmetry, may be induced by
$vev$s of the scalar components of superfields $\hat{Y}$ that enter the
superpotential through,
\begin{equation}
\hat{W}\ni \frac{\hat{Y}_1}{M_P}\hat{L}\hat{L}\hat{E}^c + 
\frac{\hat{Y}_2}{M_P}\hat{Q}\hat{L}\hat{D}^c.
\label{eq:hig}
\end{equation}
This would require $\langle Y_i\rangle \sim 10^{15}$~GeV, four orders of
magnitude larger than the scale $10^{11}$~GeV for the $vev$s of the
fields neeeded to solve the $\mu$ problem. While the existence of such
fields cannot be logically excluded, since they are not needed for
anything, we may consistently assume that these are absent. In this
case, we may expect that $\lambda, \ \lambda'\sim
\frac{\Lambda}{M_P}\sim 10^{-7}$ (or even smaller if the symmetry requires
higher powers of the MSSM singlet field). The contributions to active neutrino
masses from these couplings is then completely negligible, at least as
far as their measured oscillation parameters are concerned.

We will see below that the field $\hat{Z}^\prime$ that we introduce
along with $\hat{Z}$ to solve the $\mu$ and $b_\mu$ problems,
simultaneously plays the role of the
field $\hat{X}$ that sets the mass scale in the active neutrino
sector. Radiative corrections then allow us to accommodate (though not
explain) the required pattern of neutrino masses and mixing angles.

\section{The complete model}\label{sec:complete}

We have just seen that in addition to a hidden sector superpotential
$\hat{W}_0$ that we need 
to break supersymmetry at an intermediate scale $\Lambda
\sim 10^{11}$~GeV, we have to introduce new superfields that we will
call $\hat{X}_1$ and $\hat{X_2}$ that acquire $vev$s $\sim\Lambda$ for
their scalar components, and  $\sim {\Lambda^3\over M_P}$ for
their auxiliary components. The field $\hat{X}_2$ plays a dual role in
that it not only induces the SUSY breaking $vev$ for ${\cal F}_{X_1}$
that we need for the Kim-Nilles mechanism \cite{KNM}, but also sets the mass
scale for active neutrinos. We begin by exhibiting a model that leads to
this required pattern of $vev$s for the fields $\hat{S}$ and
  $\hat{X}_{1,2}$. 

\subsection{Dynamical origin of the vaccum expectation values}
\label{sec:D}

We begin by introducing the superpotential,
\begin{equation}
\hat{W} = \frac{\Lambda^2}{M_P^2}[M_P^2\hat{S}+\alpha M_P\hat{S}^2
+\beta \hat{S}^3 + 
\gamma M_P^3] + \frac{\kappa}{M_P}\hat{X}_1^3\hat{X}_2 +\cdots.
	\label{compsup}
\end{equation}
The first term is just $\hat{W}_0$ that we have introduced earlier. The
second term shows the lowest dimensional interaction involving the
fields $\hat{X}_{1,2}$ (that we introduce to dynamically generate the
$\mu$ and $b_\mu$ parameters as described in Sec.~\ref{subsec:mu})
invariant under the $U(1)'$ gauge symmetry. The corresponding coupling
constant $\kappa\sim 1$. We will see below that this term is essential
in that if $\kappa=0$, we will have only supersymmetric solutions. The
ellipses include superpotential couplings of $\hat{X}_1$ to the fields
$\hat{H}_{u,d}$, of $\hat{X_2}$ to $\hat{L}_i$ and $\hat{H}_u$ and
$\hat{W}_{\rm MSSM}$ that are unimportant for our analysis of the $vev$s
of $\hat{S}$ and $\hat{X}_{1,2}$. The K\"ahler potential takes the
minimal form,
\begin{equation}
\hat{K} = \hat{S}^\dagger \hat{S} + \hat{X}_1^\dagger \hat{X}_1 
+ \hat{X}_2^\dagger \hat{X}_2 +\hat{\Phi}_i^\dagger\hat{\Phi}_i +\cdots,
	\label{compkah} 
\end{equation}
consistent with the assumed symmetries. The ellipses denote
higher dimensional terms such as
$\frac{(\hat{S}+\hat{S}^\dagger)^n}
{M_P^n}\hat{X}_i^\dagger \hat{X}_i$, as well as similar terms involving
MSSM superfields that are consistent with the assumed symmetries. These
higher dimensional terms are undoubtedly present, but will only give
${\cal O}(1)$
corrections to the
solutions that we will obtain for the $vev$s, that do not qualitatively 
change the
different scales that we obtained by 
our analysis. Finally, to obtain weak scale masses for gauginos, we
choose the gauge kinetic function to be given by (\ref{eq:WKf}). Again,
higher powers of ${\hat{S}\over M_P}$ that may be present in the gauge
kinetic function will not qualitatively alter our solution. 

To facilitate our calculation of the $vev$s for the scalar components of
$\hat{S}$ and $\hat{X}_{1,2}$, we evaluate the relevant portion of the
scalar potential for our model by substituting the superpotential
(\ref{compsup}) along with our choice of the K\"ahler potential and the
gauge kinetic function into (\ref{eq:sugsca}) to obtain, 
\begin{eqnarray}
V &=& \Lambda^4\left[V_0(\alpha,\beta,\gamma,\frac{S}{M_P})+\frac{\Lambda^2}
{M_P^2}V_1(\alpha,\beta,\gamma,\kappa,\frac{S}{M_P},\frac{X_1}{\Lambda},
\frac{X_2}{\Lambda})\right.
	\nonumber \\
&&+\left.\frac{\Lambda^4}{M_P^4}V_2(\alpha,\beta,\gamma,\kappa,\frac{S}{M_P},
\frac{X_1}{\Lambda},\frac{X_2}{\Lambda})+\cdots \right]e^{K/M_P^2}
\nonumber \\
&&+\frac{g^{\prime 2}}{2}\Lambda^4\left(1+f\frac{S}{M_P}\right)^{-1}
\left[{\cal X}_1
\frac{X_1X_1^*}{\Lambda^2} + {\cal X}_2
\frac{X_2X_2^*}{\Lambda^2}\right]^2,
\label{eq:scaV}
\end{eqnarray}
where $K$ (without the caret) is the value of the K\"ahler potential,
with the superfields replaced by their scalar components.
For simplicity, we have taken
$\alpha,\beta,\gamma,\kappa$ to be real parameters. Also, $g^\prime$
that appears in the $D$-term contribution to the potential  
is the gauge coupling strength for the new U(1)$^\prime$ group.  Finally, 
${\cal X}_{1,2}$ denote the $U(1)'$ charges of the fields
$\hat{X}_{1,2}$: these evidently must satisfy $3{\cal X}_1=-{\cal X}_2$
in order for the superpotential to be $U(1)'$ invariant.

Keeping in mind that our goal is to show that this potential allows
(classical) minima with $\langle S\rangle \sim M_P$, $\langle
X_{1,2}\rangle \sim \Lambda$ with $\langle \Phi_i \rangle \ll \Lambda$,
we have written the scalar potential in terms of appropriately scaled
fields ${S\over M_P}$ and ${X_{1,2}\over \Lambda}$, and left out terms
in the observable fields $\Phi_i$ in (\ref{eq:scaV}). Although we have
no dynamical argument for selecting this vacuum solution, it is clearly
the only one that can lead to a viable phenomenology.  The dimensionless
functions
$V_0,V_1,V_2$ and $K/M_P^2$ are then all ${\cal O}$(1) and the scalar
potential itself is ${\cal O}(\Lambda^4)$.  Indeed, (\ref{eq:scaV}) is
an expansion of the scalar potential with each successive term being
smaller in magnitude by a factor ${\Lambda^2\over M_P^2}$. The ellipses
denote yet higher order terms.  The observable sector fields (that we
have not written) would enter via the functions $V_2$,
$V_3$, $\cdots$, and also via suppressed terms in the square parenthesis
in the last term of (\ref{eq:scaV}). Fortunately, these terms only result in
tiny corrections to the $vev$s of the scalar components of $\hat{S}$ and
$\hat{X}_{1,2}$, and can be neglected in our analysis. The functions
$V_0$ and $V_1$ are given by,
\begin{eqnarray}
V_0(\alpha,\beta,\gamma,\frac{S}{M_P}) &=& f_sf_s^* - 3W_sW_s^*
	\nonumber \\
V_1(\alpha,\beta,\gamma,\kappa,\frac{S}{M_P},\frac{X_1}{\Lambda},
\frac{X_2}{\Lambda}) &=&
(f_s^*\frac{S^*}{M_P}-3W_s^*)\kappa\frac{X_1^3}{\Lambda^3}\frac{X_2}{\Lambda}
+ {\rm c.c.}
	\nonumber \\
&&+\left(3\kappa\frac{X_1^2}{\Lambda^2}\frac{X_2}{\Lambda} + W_s\frac{X_1^*}
{\Lambda}\right)
\left(3\kappa\frac{X_1^2}{\Lambda^2}\frac{X_2}{\Lambda} + W_s\frac{X_1^*}
{\Lambda}\right)^*
	\nonumber \\
&&+\left(\kappa\frac{X_1^3}{\Lambda^3} + W_s\frac{X_2^*}
{\Lambda}\right)\left(\kappa\frac{X_1^3}{\Lambda^3} + W_s\frac{X_2^*}
{\Lambda}\right)^*,
\end{eqnarray}
where
\begin{eqnarray}
f_s &=& {\cal F}_S/\Lambda^2, \ \ \ {\rm with}
	\nonumber \\
{\cal F}_S &=& \frac{\partial W_{S}}{\partial S} + \frac{W_{S}}{M_P^2}
\frac{\partial K}{\partial S}, \\
W_s &=& \frac{1}{M_P^3}[M_P^2S+\alpha M_PS^2+\beta S^3 + 
\gamma M_P^3].\nonumber
\end{eqnarray}

In addition to the extremization conditions,
\begin{eqnarray}
&&\frac{\partial V}{\partial (S/M_P)} = \frac{\partial V}
{\partial (X_1/\Lambda)} = \frac{\partial V}{\partial (X_2/\Lambda)} = 0,
	\nonumber \\
\label{eq:mincon}
\end{eqnarray}
 for the fields $S$ and $X_{1,2}$,
we require that 
\begin{equation}
\langle V\rangle = 0,\quad
\langle {\cal F}_S\rangle \neq 0,
\label{eq:susycon}
\end{equation}
so that the cosmological constant vanishes (to the order that we are
evaluating it) and that supersymmetry is broken. We further require that
there are no tachyonic directions after we have shifted the fields by
their vacuum expectation values; {\it i.e} the squared scalar mass
parameters are non-negative. 

We can satisfy  these conditions order-by-order
in powers of $\Lambda^2$. Specifically, we show that for a given choice
of $\langle S\rangle \sim M_P$ and $\langle X_{1,2}\rangle \sim \Lambda$, it is
possible to choose the model parameters $\alpha, \ \beta, \ \gamma$ all 
${\cal O}(1)$, so that  (\ref{eq:mincon}) and (\ref{eq:susycon}) are
satisfied. Toward this end, we write,
\begin{eqnarray}
\frac{\langle S\rangle}{M_P} &=& a_0 + a_1\frac{\Lambda^2}{M_P^2}
+a_2\frac{\Lambda^4}{M_P^4} + \cdots,
\nonumber \\
\frac{\langle X_1\rangle}{\Lambda} &=& b_0 + b_1\frac{\Lambda^2}{M_P^2}
+b_2\frac{\Lambda^4}{M_P^4} + \cdots,
\nonumber \\
\frac{\langle X_2\rangle}{\Lambda} &=& c_0 + c_1\frac{\Lambda^2}{M_P^2}
+c_2\frac{\Lambda^4}{M_P^4} + \cdots,
\nonumber \\
\alpha &=& \alpha_0 +\alpha_1\frac{\Lambda^2}{M_P^2}
+\alpha_2\frac{\Lambda^4}{M_P^4} + \cdots,
\nonumber \\
\beta &=& \beta_0 +\beta_1\frac{\Lambda^2}{M_P^2}
+\beta_2\frac{\Lambda^4}{M_P^4} + \cdots,
\nonumber \\
\gamma &=& \gamma_0 +\gamma_1\frac{\Lambda^2}{M_P^2}
+\gamma_2\frac{\Lambda^4}{M_P^4} + \cdots,
\label{eq:par}
\end{eqnarray}
where the coefficients
$a_\bullet$'s, $b_\bullet$'s, $\cdots$ $\gamma_\bullet$'s are all
are ${\cal O}$(1). If we work to leading order, {\it i.e.}
drop all terms ${\cal O}(\Lambda^4\times \frac{\Lambda^2}{M_P^2})$, it
is clear that we must separately minimize the first and last terms of
(\ref{eq:scaV}), and also satisfy (\ref{eq:susycon}). Minimization of
the $D$-term contribution to the scalar potential then gives us (we take
the $vev$s to be real),
\begin{equation}
{\cal X}_1
\frac{\langle X_1\rangle^2}{\Lambda^2} + {\cal X}_2
\frac{\langle X_2\rangle^2}{\Lambda^2} = 0, \ \ \ {\rm or} \ 3{\cal X}_1=-{\cal X}_2.
\label{eq:D}
\end{equation}
Extremization with respect to $S$ and the vanishing of the cosmological
constant then give, 
\begin{equation}
(\langle f_s\rangle^2-3\langle W_s\rangle^2)e^{K/M_P^2}
= 0,\quad
\frac{\partial (V_0e^{K/M_P^2})}{\partial (S/M_P)} = 0.
\label{eq:lead}
\end{equation}
It is clear that for given values of $a_0$, $b_0$ and
$c_0$, we can always
choose two of the three parameters $\alpha_0$, $\beta_0$ and $\gamma_0$
to satisfy these conditions.

While the ratio ${\langle X_1\rangle}\over {\langle X_2\rangle}$ is
fixed even at leading order, the scale of $\langle X_{1,2}\rangle$ is
still arbitrary. This degeneracy of the potential is removed
once we take the terms $\sim \Lambda^6/M_P^2$ that appear in $V_1$ into
account. The extremization conditions then give us,
\begin{equation}
\left(\frac{\langle X_2\rangle}{M_P}\right)
\left[81\left(\frac{\kappa\langle X_2\rangle^2}{\Lambda^2}\right)^2+
3\sqrt{3}\left(\langle f_s\rangle\frac{\langle S\rangle}{M_P}+
\langle W_s\rangle\right)
\frac{\kappa\langle X_2\rangle^2}{\Lambda^2}+\langle W_s\rangle^2\right] = 0.
\label{eq:X}
\end{equation}
We obtain non-vanishing values of $\langle X_2\rangle$ if the second
factor vanishes. 
Substituting $f_s=\sqrt{3}W_s$ from the first equation in (\ref{eq:lead}),
we see that we obtain real solutions for $\langle X_2\rangle$
provided,
\begin{equation}
\left|\frac{\langle S\rangle}{M_P}\right|\geq 2-\frac{1}{\sqrt{3}}\approx 1.42.
\end{equation}

The reader may be concerned that we are assuming the effective theory
assumed to be valid below the Planck scale to derive Planck scale $vev$s
for which yet higher powers of $\hat{S}$ in $\hat{W}_0$ (these are not
forbidden by any symmetry) may be important. Moreover, the
superpotential could also include terms such as
$\frac{\hat{S}^n}{M_P^n}\times\frac{\hat{X}_1^3\hat{X}_2}{M_P}$ (as well
as corresponding terms involving MSSM superfields) that would, for large
enough $n$, destabilize $\kappa$ in (\ref{compsup}) from its value of
${\cal O}(1)$, if $\langle S \rangle > M_P$.  The point, however, is
that none of our conclusions from this point on will depend on the
choice of $\hat{W}_0$. The rest of our analysis (which determines the
observable particle physics) would be qualitatively unchanged even if
$\hat{W}_0$ is not a polynomial and radiative corrections are included,
as long as $\langle S\rangle \sim M_P$. Thus, while the precise value of
the $vev$ is not trustworthy, our conclusions about various scales
arrived at using the fact that $\langle S\rangle \sim M_P$ are
reliable. Put somewhat differently, we have assumed that the potential
of the high energy theory has a local minimum (our vaccuum), with a SUSY
breaking scale $\Lambda \ll M_P$ and a cosmological constant that is
fine-tuned to be (almost) zero, sufficiently separated from other
minima. An expansion about this minimum, (rather than about $S=0$) then
leads to an effective field theory in which the higher dimensional
operators will indeed all be suppressed by corresponding powers of
$M_P$.

We now have to check whether the extremum that we have obtained is
indeed a local minimum. Toward this end, in Table~\ref{tab:sol} we give
an illustrative example of a solution\footnote{There will be a
corresponding solution for a negative value of $\langle S\rangle$ with
the same spectrum and the same value of $\beta_0$, but with the signs of
$\alpha_0$, $\gamma_0$ and $\kappa b_0^2$ flipped in.} for
$\langle S\rangle = 1.5 M_P$.  The spontaneous breakdown of the $U(1)'$
gives a massless would-be Goldstone boson (a linear combination of the
imaginary components of the $X_{1,2}$ fields) that makes the $U(1)'$
gauge field massive via the Higgs mechanism. A corresponding combination
of the real parts of $X_1$ and $X_2$ acquire a large mass $m_{Xh}\sim
\Lambda$, the precise value depends on the details including
parameters in the gauge kinetic function. The corresponding orthogonal
combination $X_l$ ($h$ and $l$ here denote heavy and light) as well as
the non-Goldstone combination of $X_{1I}$ and $X_{2I}$ get masses $\sim
(2-20)\Lambda^2/M_P$ from the interactions contained in the $V_1$ part
of the scalar potential (\ref{eq:scaV}).  The real and imaginary parts
of the singlet $S$ also acquire TeV scale masses. The
positive values of the squared mass parameters indeed demonstrate that
we have a true minimum.  We remark that for solutions at the lower
extreme $\left|\frac{\langle S\rangle}{M_P}\right|=
2-\frac{1}{\sqrt{3}}$, $m_{Xl}^2=0$, so that the state $X_l$ which may
be very light, could have implications for Higgs physics as well as for
cosmology.
\begin{table}
\begin{tabular}[t]{||c|c|c|c|c|c|c|c|c|c|c||}\hline
$a_0$ & $\gamma_0$ & $\kappa b_0^2$
& $\alpha_0$ & $\beta_0$ & $\langle{\cal F}_S\rangle$ & $m_{SR}^2$
& $m_{SI}^2$ & $m_{Xh}^2$ & $m_{Xl}^2$ & $m_{XI}^2$\\
 & & & & & $(\Lambda^2)$ & $(\frac{\Lambda^4}{M_P^2})$ & $(\frac{\Lambda^4}
{M_P^2})$ & $(\Lambda^2)$ & $(\frac{\Lambda^4}{M_P^2})$ & $(\frac{\Lambda^4}
{M_P^2})$ \\\hline
1.5 & -.35 & -.053 & -.44 & .059 & .63 & 5.6 & 14.4 & $\sim$1 & 7.4 & 54.8 \\
1.5 & -.2 & -.106 & -.21 & -.029 & 1.26 & 15.5 & 64.8 & $\sim$1 & 29.7 & 219.4 
\\
1.5 & -.1 & -.14 & -.059 & -.088 & 1.67 & 8.25 & 134 & $\sim$1 & 52.7 & 390 \\
1.5 & -.05 & -.16 & .017 & -.118 & 1.88 & 2.49 & 177.6 & $\sim$1 & 66.7 & 493.6
\\\hline
\end{tabular}
\caption{Sample solutions with zero vaccum energy and local minima along
with the leading values of the parameters in (\ref{eq:par}). We have
also given SUSY breaking parameter ${\cal F}_S$ in units of $\Lambda^2$
and also the squared masses of the various scalar fields in units of
either $\Lambda^2$ or $(\Lambda^2/M_P)^2$ as appropriate. As discussed
in the text, the corresonding solution for $\langle S \rangle=-1.5M_P$
has an identical spectrum.}
\label{tab:sol}
\end{table}


\subsection{Anomalies} \label{subsec:anomalies}
Since our solutions to the $\mu$ and $b_\mu$ problem,
as well as the mechanism for neutrino masses discussed in Sec.~\ref{sec:C},
both require us to extend the gauge
group, we need to ensure that the associated anomalies cancel. We will
see shortly that this will require us to introduce new fields (some of
which are charged under the MSSM gauge group, $SU(3)_C\times
SU(2)_L\times U(1)_Y$) that may manifest
themselves as exotic particles at the multi-TeV scale \cite{anomaly,ma2}.

We note that the invariance of the usual quark Yukawa
coupling terms in the superpotential require that 
\begin{equation}
2{\cal Q} + {\cal U}^c + {\cal D}^c + {\cal H}_u + {\cal H}_d=0,
\label{eq:fn}\end{equation}
with the understanding that the calligraphic symbol for the field 
denotes its $U(1)'$ charge. Since an important role of the $U(1)'$ field
was to forbid the $\mu$ term, we know that ${\cal H}_u + {\cal H}_d
\not=0$, from which we infer that
\begin{equation}
2{\cal Q} + {\cal U}^c + {\cal D}^c \not=0.
\label{eq:yukineq}
\end{equation}

We commence our discussion of the anomalies by observing that the new
fields $\hat{S}$ and $\hat{X}_{1,2}$ are $SU(3)_C\times SU(2)_L\times
U(1)_Y$ singlets, and so do not spoil the anomaly cancellation of the
MSSM. We thus need to focus only on the mixed anomalies involving the
MSSM gauge group or gravity and the new $U(1)'$ gauge group. We begin
with the cancellation of the $[SU(3)_C]^2\times U(1)'$ anomaly which, with
the field content that we have up to this point, would require $$2{\cal
Q} + {\cal U}^c + {\cal D}^c=0,$$ in direct contradiction with
(\ref{eq:yukineq}). We are thus led to introduce new $SU(2)_L$ singlet,
colour triplet fields $\hat{K}$ and $\hat{K}^\prime$ in the ${\bf 3}$
and ${\bf 3}^*$ representation of $SU(3)_C$ with $U(1)'$ quantum numbers
${\cal K}$ and ${\cal K}^\prime$, respectively, and weak hypercharge
$y(K)=-y(K^\prime)$. Since $\hat{K}$ and $\hat{K}^\prime$ are in
conjugate representations of the MSSM gauge group, their introduction
does not affect the cancellation of the $SU(3)_C\times SU(2)_L\times
U(1)_Y$ and the mixed $[gravity]^2-U(1)_Y$ anomalies. We choose their
$U(1)'$ charges to cancel the mixed $[SU(3)_C]^2\times U(1)'$
anomaly. However, we also need to ensure that the coloured superfields
acquire a mass. The simplest way to do so is to introduce a
superpotential coupling $$\hat{W}\ni
\frac{\hat{X}_1^2\hat{K}\hat{K}^\prime}{M_P}$$ which is consistent with
the cancellation of the anomaly {\it provided we introduce one pair of
$\hat{K}$ and $\hat{K}^\prime$ for each matter family.} To understand
the reason for this, we first note that the cancellation of the
$[SU(3)_C]^2\times U(1)'$ anomaly then requires that,
\begin{eqnarray}
&[SU(3)_C]^2-U(1)^\prime &:\quad 3(2{\cal Q} + {\cal U}^c + {\cal D}^c) +
\sum_{i=1}^{n_K}\left({\cal K}_i + {\cal K}^\prime_i\right) = 0,
\label{eq:A1}
\end{eqnarray}	
where the factor 3 on the first term on the right-hand-side arises
because there are three quark families, and the index $i = 1-n_K$ counts
different sets of $\hat{K}$ and $\hat{K}^\prime$ fields (with the same
$U(1)'$ quantum numbers).  Since our solution to the $\mu$ and $b_\mu$
problems requires that ${\cal H}_u+{\cal H}_d=-2{\cal X}_1$, we infer
from (\ref{eq:A1}) that the lowest dimensionality $U(1)'$-invariant
superpotential operator that can give the new coloured fields a mass
is $\hat{X}_1^{\frac{6}{n_K}} \hat{K}\hat{K}^\prime$, where $n_K = 1,3$
or 6, is the number of pairs of these coloured fields. The corresponding
mass for these fields is $\sim {{\Lambda^{\frac{6}{n_K}}}\over
{M_P^{{\frac{6}{n_K}}-1}}}$, which for $\Lambda\sim 10^{11}$~GeV is
unacceptably small for $n_K=1$ but leads to the interesting prediction
of new coloured states at the multi-TeV scale if $n_K=3$ is the number
of quark generations.

The superpotential, for quarks, leptons, Higgs and the new superfields
that various considerations have led us to introduce, must include
\begin{equation}
\hat{W} \ni {\bf Y}^u_{ij}\hat{Q}_i\hat{U}^c_j\hat{H}_u +
{\bf Y}^d_{ij}\hat{Q}_i\hat{D}^c_j\hat{H}_d +
{\bf Y}^e_{ij}\hat{L}_i\hat{E}^c_j\hat{H}_d + a\frac{\hat{X}_1^2}{M_P}\hat{H}_u
\hat{H}_d + a'\frac{\hat{X}_1^3\hat{X}_2}{M_P} + h_i\frac{\hat{X}_2^3}{M_P^2}\hat{L}_i\hat{H}_u +
a^{''}_i\frac{\hat{X}_1^2}{M_P}\hat{K}_i\hat{K}^\prime_i + \cdots,
\label{eq:Wob}
\end{equation}
where ${\bf Y}$'s are the usual quark and lepton Yukawa coupling
matrices, and $a$, $a'$, $a^{''}_i$ and 
$h_i$ are dimensionless coupling
constants assumed to be  ${\cal O}$(1), and a sum over the indices $i$
and $j$ is implied. The penultimate term leads to
neutrino masses as discussed in Sec.~\ref{sec:C}.   
The last term in (\ref{eq:Wob}), that gives supersymmetric masses to the
$\hat{K}_i$ and $\hat{K}^\prime_i$ has been written in the diagonal
basis for these fields. The ellipses include yet higher dimensional
operators in the superpotential that would be allowed by the gauge
symmetry but are  of  no relevance to us,
along with the dynamics of the SUSY-breaking singlet $\hat{S}$ that we have
already discussed. 

The conditions for the cancellation of the remaining gauge and mixed
gauge-gravity anomalies that supplement (\ref{eq:A1}) above read,
\begin{eqnarray}
&[SU(2)_L]^2-U(1)^\prime &:\quad 9{\cal Q} + 3{\cal L} + {\cal H}_u +
{\cal H}_d = 0,
\label{eq:A2}
	\\
&[U(1)_Y]^2-U(1)^\prime&:\quad 2{\cal Q} + 16{\cal U}^c + 4{\cal D}^c +3
(2{\cal L}+4{\cal E}^c)
+ 2{\cal H}_d + 2{\cal H}_u 
	\nonumber \\
&&+ 3\sum_{i=1}^3 \left(y(K_i)^2{\cal K}_i+y(K^\prime_i)^2
{\cal K}^\prime_i\right) = 0,
\label{eq:A3}
	\\
&[\rm gravity]^2-U(1)^\prime&:\quad 9(2{\cal Q}+{\cal U}^c+{\cal D}^c) +
3(2{\cal L}+{\cal E}^c)
+ 2({\cal H}_d+{\cal H}_u) + N_1{\cal X}_1 + N_2{\cal X}_2 
	\nonumber \\
&&+ N_3{\cal Y}_1 + N_4{\cal Y}_2 +
3\sum_{i=1}^3\left({\cal K}_i+{\cal K}^\prime_i\right) = 0,
\label{eq:A4}
	\\
&U(1)_Y-[U(1)^\prime]^2&:\quad 3(2{\cal Q}^2 - 4{\cal U}^{c2} +
2{\cal D}^{c2}) + 3(-2{\cal L}^2
+2{\cal E}^{c2}) + 2(-{\cal H}^2_d+{\cal H}^2_u) 
	\nonumber \\
&&+ 3\sum_{i=1}^3
\left(y(K_i){\cal K}^2_i+y(K^\prime_i){\cal K}^{\prime 2}_i\right) = 0, \ \ \ {\rm and}
\label{eq:A5}
	\\
&[U(1)^\prime]^3&:\quad 9(2{\cal Q}^3+{\cal U}^{c3}+{\cal D}^{c3}) +
3(2{\cal L}^3+{\cal E}^{c3})
+ 2({\cal H}^3_d+{\cal H}^3_u) + N_1{\cal X}^3_1 + N_2{\cal X}^3_2 
	\nonumber \\
&&+ N_3{\cal Y}^3_1 + N_4{\cal Y}^3_2 +
3\sum_{i=1}^3\left({\cal K}^3_i+{\cal K}^{\prime 3}_i\right) = 0.
\label{eq:A6}
\end{eqnarray}
The reader will have to notice that we have included two additional
$SU(3)_C\times
SU(2)_L\times U(1)_Y$ singlet superfields $\hat{Y}_1$ and $\hat{Y_2}$
with non-trivial $U(1)'$ charges that do not alter anomalies
involving any SM gauge boson in our analysis. As we will see below,
their inclusion is only necessary if we want to obtain rational values
for all $U(1)'$ charges. We have also allowed for several copies
($N_1,\ N_2,\ N_3,\ N_4$) of the SM singlet superfields ($\hat{X}_1, \
\hat{X}_2, \ \hat{Y}_1, \ \hat{Y}_2$, respectively). We must thus understand
that the couplings $a$, $a'$, $a^{''}$ as well as the coupling $h_i$
that appear in (\ref{eq:Wob}) carry an extra index that specifies just
which one of these multiple singlet fields we are referring to. 

The various $U(1)'$ charges are, of course, not independent since the
corresponding invariance of $\hat{W}$ in ({\ref{eq:Wob}) requires that,
\begin{eqnarray}
&&{\cal Q} + {\cal U}^c + {\cal H}_u = 0,
\label{eq:up}
	\\
&&{\cal Q} + {\cal D}^c + {\cal H}_d = 0,
\label{eq:dow}
	\\
&&{\cal L} + {\cal E}^c + {\cal H}_d = 0,
\label{eq:cha}
	\\
&&2{\cal X}_1 + {\cal H}_d + {\cal H}_u = 0,
\label{eq:XHH}
	\\
&&3{\cal X}_2 + {\cal L} + {\cal H}_u = 0,
\label{eq:XLH}
	\\
&&2{\cal X}_1 + {\cal K}_i + {\cal K}^\prime_i = 0,
\label{eq:XKK} \\
&&{\cal X}_2 +3{\cal X}_1 =0. \label{eq:XX}
\end{eqnarray}
Notice that we can eliminate ${\cal H}_u$, ${\cal H}_d$ and ${\cal X}_1$
from (\ref{eq:up}), (\ref{eq:dow}), (\ref{eq:XHH}) and (\ref{eq:XKK}) to
obtain (\ref{eq:A1}). The other equations are all independent and, along
with the ratio ${\cal Y}_1/{\cal X}_1$ which is fixed to obtain rational
values of $U(1)'$ charges as discussed below, can be used to write the
$U(1)'$ charges of the seventeen fields (up to discrete quadratic or
cubic ambiguities) in terms of the charges of any two fields which we
will take to be ${\cal X}_1$ and ${\cal L}$. Our aim is to display one
such solution explicitly.

Toward this end, we remark that by considering the linear combination
$$\frac{1}{2}\times(29)+(28)-8\times (33)-2\times (34) -6\times (35),$$
of equations (\ref{eq:A2}), (\ref{eq:A3}), (\ref{eq:up}), (\ref{eq:dow})
and (\ref{eq:cha}), 
and  noting that ${\cal H}_u + {\cal H}_d=-2{\cal X}_1 = {\cal
  K}_i+{\cal K}_i^\prime$, we obtain (since ${\cal X}_1\not=0$)
\begin{equation}
\sum_{i=1}^3y(\hat{K}_i)^2 = 4.
\label{eq:hyp}
\end{equation}
This choice of weak hypercharges is necessary to
guarantee  the cancellation of the $[U(1)_Y]^2-U(1)^\prime$ anomaly.

Next, in (\ref{eq:A4}) for the cancellation of the $[{\rm
gravity}]^2-U(1)^\prime$ anomaly, we first note that the first and last
terms sum to zero. Then, using (\ref{eq:cha})--(\ref{eq:XLH}) together
with ${\cal X}_2=-3{\cal X}_1$, we find that,
\begin{equation}
(N_1-3N_2+29){\cal X}_1 + N_3{\cal Y}_1 +N_4{\cal Y}_2=0,
\label{eq:nouse}
\end{equation}
which shows why multiple copies of some of these MSSM singlet fields are
 necessary. 
 A simple solution is given by,\footnote{We emphasize that while
 (\ref{eq:hyp}) and (\ref{eq:nouse}) must always be satisfied, from here
 on, our focus will be to exhibit an  explicit solution
of the anomaly constraints. Other solutions may also be possible. We
 have checked though that it is not possible to satisfy the anomaly
 equations if $N_1=N_2=N_3=N_4=1$.}
\begin{eqnarray}
&&N_1 = 1,\quad N_2 = 10,\quad {\cal X}_2 = -3{\cal X}_1,
	\nonumber \\
&&N_3 = 1,\quad N_4 = 2,\quad {\cal Y}_1 = -2{\cal Y}_2.
\label{eq:NXY}
\end{eqnarray}

We now turn to the last two anomaly constraints, the quadratic and cubic
equations, (\ref{eq:A5}) and ({\ref{eq:A6}), for the $U(1)'$
  charges. These depend explicitly on the weak hypercharges of the
  fields $\hat{K}_i$ which, as we have seen, satisfy
  (\ref{eq:hyp}).  Of the many possible solutions, we first make the  
simple choice $y(\hat{K}_i) = (2,0,0)$ which obviously leads to
integrally charged, coloured particles. To find a solution, we first use
(\ref{eq:up})--(\ref{eq:XX}) together with (\ref{eq:A2}) to eliminate
all $U(1)'$ charges in terms of ${\cal X}_1$, ${\cal L}$ and ${\cal
  K}_1$, and then plug these into (\ref{eq:A5}) to obtain (note that
${\cal K}_2$ and ${\cal K}_3$ drop out because the corresponding weak
hypercharges are chosen to be zero),
\begin{equation}
\frac{8}{3}{\cal X}_1(-9{\cal L}+56{\cal X}_1) - 12{\cal X}_1{\cal K}_1 = 0.
\label{eq:k1}
\end{equation}

Finally, we turn to the $[U(1)^\prime]^3$ anomaly equation
(\ref{eq:A6}). Writing all but ${\cal K}_i$ and ${\cal Y}_1$ in terms of
${\cal L}$ and ${\cal X}_1$ (remember that ${\cal Y}_1=-2{\cal Y}_2$),
this reduces to,
\begin{equation}
\sum_{i=1}^3{\cal K}_i^2+2{\cal X}_1{\cal K}_i = \frac{1}{18{\cal X}_1}
\{72{\cal L}^2{\cal X}_1-968{\cal L}{\cal X}_1^2+\frac{19970}{3}{\cal X}_1^3
+\frac{3}{4}{\cal Y}_1^3\}.
\end{equation}
We now eliminate ${\cal K}_1$ using  (\ref{eq:k1}) and find,
\begin{equation}
({\cal K}_2+{\cal X}_1)^2 + ({\cal K}_3+{\cal X}_1)^2 =
\left(\frac{{\cal X}_1}{9}\right)^2\left[15557+\left(\frac{3{\cal Y}_1}
{2{\cal X}_1}\right)^3\right].
\label{eq:K2K3}
\end{equation}
Remarkably, ${\cal L}$ does not appear in this equation, which has
solutions with {\it rational values} of $U(1)'$ charges for
$\frac{3{\cal Y}_1}{2{\cal X}_1}= -1$, $-3$, $-4$, $-6$,
$-10,\cdots$.\footnote{We focus on negative values only to limit the
magnitudes of the $U(1)'$ charges ${\cal K}_{2,3}$.}  We now see the
role of the $\hat{Y}_{1,2}$ fields. Without these, the anomaly
constraints would be satisfied, only for irrational values of the
$U(1)'$ charges, precluding the possibility of embedding the model into
a grand unified framework with a simple gauge group.\footnote{This is
also the reason that we do not include a kinetic mixing between the
$U(1)_Y$ and $U(1)'$ gauge particles. The main low energy effect of such
a mixing would be to alter the usual MSSM $D$-term contribution to the
scalar mass parameters, which would now depend not only on $M_Z$ and
$\tan\beta$, but also on an additional parameter characterizing the
mixing \cite{DKM}. }

We note that it is also possible to satisfy the anomaly equations with
$y(\hat{K}_i) = \left(\frac{2}{3}, \frac{4}{3},\frac{4}{3}\right)$ which
leads to electric charges $\pm \frac{1}{3}$ or
$\pm\frac{2}{3}$ for the coloured $\hat{K}_i$ and $\hat{K}^\prime_i$
fields. Again solutions are possible for $\frac{3{\cal Y}_1}{2{\cal
X}_1}= -1$, $-3$, $-4$, $-6$, $-10,\cdots$. In this case, the reader may
suppose that the new coloured particles may mix (upon spontaneous
breaking of  $U(1)'$) with the usual $SU(2)$-singlet quarks and squarks
with the same spin and electric charge. However, as we will see below,
there are simple cases where such mixing is precluded.

To recapitulate, the requirement of anomaly cancellation forces us to
introduce three pairs of coloured fields $\hat{K}_i$  and
$\hat{K}_i^\prime$,  with either integral or fractional electric charges,
with masses at the TeV scale. We also require SM
singlet superfields $\hat{X}_{1,2}$ and $\hat{Y}_{1,2}$, with ten copies
of $\hat{X_2}$ and two of
$\hat{Y}_2$. The $\hat{Y}_{1,2}$ fields are necessary only if we insist
on rational values of $U(1)'$ charges. There are many solutions to the  
anomaly equations. Here, we exhibit an explicit solution with  $\frac{3{\cal
    Y}_1}{2{\cal X}_1}= -22$, for which the new
$U(1)'$ charges are fixed by the corresponding charges of $\hat{X}_1$
and $\hat{L}_i$ fields by,
\begin{eqnarray}
&&{\cal X}_2 = -3{\cal X}_1,\quad {\cal Y}_1 = -\frac{44}{3}{\cal X}_1,
\quad {\cal Y}_2 = \frac{22}{3}{\cal X}_1,
	\nonumber \\
&&{\cal K}_1 = \frac{2}{9}(-9{\cal L}+56{\cal X}_1) \
		  \left[-\frac{2}{3}{\cal L} - \frac{40}{27}{\cal
		      X}_1\right], \quad
{\cal K}_2 = \frac{61}{9}{\cal X}_1 \ 
\left[-\frac{4}{3}{\cal L} + \frac{139}{27}{\cal X}_1\right], \nonumber \\
&& {\cal K}_3 = -\frac{2}{3}{\cal X}_1 \ \left[-\frac{4}{3}{\cal L} +
  \frac{358}{27}{\cal X}_1\right], \quad
	{\cal K}^\prime_i = -{\cal K}_i - 2{\cal X}_1,\quad  i=1,2,3,
\label{eq:rel}	 \\
&&{\cal Q}=(-3{\cal L}+2{\cal X}_1)/9,\quad
{\cal D}^c = -\frac{2}{3}{\cal L}+\frac{97}{9}{\cal X}_1,\quad
{\cal U}^c = \frac{4}{3}{\cal L}-\frac{83}{9}{\cal X}_1,
	\nonumber \\
&&{\cal E}^c = -2{\cal L}+11{\cal X}_1,\quad
{\cal H}_u = -{\cal L}+9{\cal X}_1,\quad
{\cal H}_d = {\cal L}-11{\cal X}_1, \nonumber
\end{eqnarray}
where the values of ${\cal K}_i$ outside (within) the square parenthesis
refer to the integrally (fractionally) charged case for the $\hat{K}_i$
fields discussed above.  For rational values of ${\cal L},{\cal X}_1$,
we obtain rational U(1)$^\prime$ charges for all the other fields. We
have checked that despite the large number of fields that we have
introduced, the $U(1)'$ gauge coupling does not blow up below $Q=M_P$ as
long as the corresponding weak scale gauge coupling is smaller than
about 0.05 [0.6] for $({\cal L}, {\cal X}_1)$ = (1,1) [(1, 0.1)], {\it
i.e.} as long as the couplings of fields such as $\hat{E}^c$,
$\hat{D}^c$, {\it etc.} that have a much larger coupling to the
$U(1)'$ gauge boson than $\hat{X}_1$, are similar in magnitude to the SM
gauge couplings.

The alert reader will have noted that since we chose the weak
hypercharges for the colour triplet fields to be positive, it is the
charged $\frac{1}{3}$ triplet $\hat{K}_1$, not the anti-triplet
$\hat{K}_1^\prime$, that has positive charge. Before closing this
discussion, we point out how the $U(1)'$ charges in (\ref{eq:rel}) would
be altered had we instead chosen the
weak hypercharge $y(\hat{K}_1)=-\frac{2}{3}$. This would have only shown
itself only in the $U(1)_Y-U(1)^{\prime 2}$ anomaly equation
(\ref{eq:A5}), which does not distinguish whether the
$\hat{K}_i^{(\prime)}$ are triplets or anti-triplets. The flip of the
sign of the weak hypercharge of $\hat{K}_1$ thus results in an
interchange of the $U(1)^\prime$ charges of $\hat{K}_1$ and
$\hat{K}_1^\prime$ from their values (in the square parenthesis) in
(\ref{eq:rel}), with the $U(1)^\prime$ charges of all other fields
remaining unchanged.

\subsection{A recap}
With the $U(1)'$ charges that we have just obtained in (\ref{eq:rel}), the
most general superpotential, invariant under the assumed symmetries, 
may be written as,
\begin{eqnarray}
W &=& \frac{\Lambda^2}{M_P^2}[M_P^2 \hat{S}+\alpha M_P \hat{S}^2+
\beta \hat{S}^3+\gamma M_P^3 ]+\sum_{a=1}^2 y_a \hat{Y}_1\hat{Y}^2_{2a} +
\sum_{b=1}^{10}\frac{\kappa_b}{M_P}\hat{X}_1^3\hat{X}_{2b} +
Y^u_{ij}\hat{Q}_i\hat{U}^c_j\hat{H}_u +	\nonumber \\
&&
Y^d_{ij}\hat{Q}_i\hat{D}^c_j\hat{H}_d +
Y^e_{ij}\hat{L}_i\hat{E}^c_j\hat{H}_d + 
\frac{\hat{X}_1^2}{M_P}\hat{H}_u\hat{H}_d + 
\sum_{b=1}^{10}h_{ib}\frac{\hat{X}_{2b}^3}{M_P^2}\hat{L}_i\hat{H}_u +
a_i^{''}\frac{\hat{X}_1^2}{M_P}\hat{K}_i\hat{K}^\prime_i+\cdots,
\label{eq:sugW}
\end{eqnarray}
where a sum over $i, j$ is implied, and $y_a$ are
couplings of ${\cal O}$(1). Here, we have explicitly shown the sums over
the MSSM singlet fields $\hat{X}_{2b}$ and $\hat{Y}_{2a}$, while the
sums over the family indices $i$ and $j$ (including for the colour
triplet superfields $\hat{K}_i$ and $\hat{K}^\prime_i$) are implied. The
ellipses denote other terms allowed by the symmetries, but suppressed by
even higher powers of $M_P$, that are irrelevant for our analysis.

It is instructive to note that trilinear $R$-parity violating operators,
$$\hat{L}_i\hat{L_j}\hat{E}^c_k, \ \hat{L}_i\hat{Q}_j\hat{D}^c_k, \ {\rm
  and} \ \hat{U}^c_i\hat{D}^c_j\hat{D}^c_k,$$
are automatically forbidden by the $U(1)'$ gauge symmetry. 
Since $2{\cal L}+{\cal E}^c = {\cal L}+{\cal Q}+ {\cal D}^c = 11{\cal
  X}_1$, the first two operators may be induced, by the spontaneous
  breaking of $U(1)'$,  via effective terms, 
$$\frac{(\hat{X}_1^\dagger)^2\hat{X}_2^3}{M_P^6}\hat{L}\hat{L}\hat{E}^c,
\frac{(\hat{X}_1^\dagger)^2\hat{X}_2^3}{M_P^6}\hat{L}\hat{Q}\hat{D}^c$$
in the K\"ahler potential which, because $\langle F_{X_1}\rangle
\sim\frac{\Lambda^3}{M_P}$ and $\langle X_{1,2}\rangle\sim \Lambda$,
leads to associated dimensionless couplings $\sim
\frac{\Lambda^7}{M_P^7}\sim 10^{-49}$ which are utterly negligible.
Since ${\cal U}^c+2{\cal D}^c=\frac{37}{3}{\cal X}_1$, the baryon-number
violating trilinear superpotential interaction
$\hat{U}^c_i\hat{D}^c_j\hat{D}^c_k$ cannot be induced.

\section{Phenomenology}\label{sec:phen}

\subsection{Proton Decay and $n \bar{n}$ oscillations}
We have just seen that the $U(1)'$ gauge symmetry that we have
introduced automatically suppresses all dimensionless baryon and lepton
number violating superpotential couplings to negligible levels. Thus,
the introduction of {\it ad hoc} global symmetries to avoid the
disastrous prediction of proton decay {\it at the weak interaction
rate} is unnecessary within our model. 

The dimension-4 superpotential operators (which generate dimension-5
terms in the interaction Lagrangian),
$$\hat{Q}\hat{Q}\hat{Q}\hat{L} \ {\rm and} \
\hat{U}^c\hat{U}^c\hat{D}^c\hat{E}^c,$$ can lead to a dangerously high
rate for proton decay \cite{prodecay}.  It is, however, easy to see that
the $U(1)'$ gauge symmetry also forbids these operators. Moreover, these
operators are not induced even after $U(1)'$ breaking because $3{\cal
Q}+{\cal L}=\frac{2}{3}{\cal X}_1$, while $2{\cal U}^c+{\cal D}^c+{\cal
E}^c=\frac{30}{9}{\cal X}_1$. The baryon-number violating (but
lepton-number conserving) operator, $\hat{Q}\hat{Q}\hat{Q}\hat{H}_d$ is
also not possible. We also note that baryon- and lepton-number violating
terms, $$\hat{Q}\hat{Q}\hat{U}^{c\dagger}\hat{E}^{c\dagger} \ {\rm and} \
\hat{Q}\hat{U}^{c\dagger}\hat{D}^{c\dagger}\hat{L}, $$ in the K\"ahler
potential, which give rise to gauge-boson-mediated proton decay in
many SUSY grand unified theories are also forbidden.\footnote{We mention
  in passing that dimension-5 lepton-number-violating 
interactions generated by the
 operators $\hat{L}\hat{L}\hat{H}_u\hat{H}_u$ or  
$\hat{Q}\hat{U}^c\hat{E}^c\hat{H}_d$ in the superpotential, or 
the operators $\hat{U}^c\hat{D}^{c\dagger}\hat{E}^c$ or
 $\hat{Q}\hat{U}^c\hat{L}^\dagger$ in the K\"ahler potential are
 allowed, but very strongly suppressed by a high power (6, for the first
 operator, and even larger for the other operators) of 
$\frac{\langle X_1 \rangle}{M_P}$ in addition to the usual factor of $M_P$.} 
Our model is thus safe from  constraints from the
non-observation of proton decay at Super-Kamiokande \cite{superK}. 

Indeed, it turns out that the proton is stable within this
framework.\footnote{We thank Christoph Luhn for pointing this out to
us.}  To see this, we first observe that the coefficients multiplying
${\cal L}$ in the $U(1)'$ charges in (\ref{eq:rel}) are proportional to
the weak hypercharges of the corresponding fields, so that changing just
the value of ${\cal L}$ amounts to just a $U(1)_Y$ transformation, under
which our Lagrangian is automatically invariant. Next, taking ${\cal
L}=\frac{2}{3}{\cal X}_1$, we observe that the $U(1)'$ charges of the
MSSM fields, normalized so that ${\cal X}_1=3$, are given by,
\begin{equation} 
({\cal Q},{\cal U}^c,{\cal D}^c,{\cal L},{\cal E}^c,{\cal H}_d,{\cal
  H}_u,{\cal X}_1)  
= (0,-25,31,2,29,-31,25,3)=(0,2,1,2,2,2,1,0) \  {\rm mod} \ 3.\label{eq:z3}
\end{equation}
Since only those products of MSSM fields ${\cal O}_{\rm MSSM}$ that are
of the form ${\cal O}_{\rm MSSM}\times X_1^{(\dagger)n}X_2^{(\dagger)m}$
(with $m$ and $n$ being integers) arise in the low energy theory even
after spontaneous breakdown of the $U(1)'$ group, we conclude that the
$U(1)'$ charge of ${\cal O}_{\rm MSSM}$ must vanish modulo 3. In other
words, a $Z_3$ subgroup of $U(1)'$ with charges of MSSM fields as in
(\ref{eq:z3}) remains as a discrete symmetry of the low energy
theory even though $U(1)'$ is spontaneously broken.\footnote{There is a
potential loophole in this argument since it is possible that the fields
$Y_{1,2}$ acquire $vev$s. Then, for instance, if ${\cal Y}_1/{\cal X}_1$
is not an integer, the $vev$ of $Y_1$ would break the $Z_3$ symmetry,
obviating our argument. However, even though a $\Delta B=1$ operator may
then be allowed in the low energy theory, we would expect that it would
have a very high dimensionality, so that the proton decay rate would
still be very suppressed by appropriate powers of $M_P$.} Indeed, this
$Z_3$ is the same as the discrete symmetry group $B_3$ first discussed
by Iba\~nez and Ross \cite{IR}. The $Z_3$ charges of MSSM fields in
(\ref{eq:z3}) coincide with the corresponding $2(B-Y)$ charges modulo 3,
so that conservation of $Z_3$ implies that the $SU(3)_C\times
SU(2)\times U(1)_Y$ invariant low energy effective theory conserves $2B$
modulo 3, or that $\Delta B=1,2$ processes are forbidden \cite{cm}. The
proton is thus stable within our framework, and further, neutron
anti-neutron oscillations are also forbidden (or at least very strongly
suppressed if $Y_1$ or $Y_2$ acquires a $Z_3$ breaking $vev$).


\subsection{Neutrino sector}

We have already seen in Sec.~\ref{sec:C} that because $\hat{X}_{2b}$
fields acquire $vevs$ $\sim \Lambda$, the penultimate term in
the superpotential (\ref{eq:sugW}) naturally results in a neutrino mass
scale $\sim 0.1$~eV. We have also seen that lepton-number-violating
trilinear couplings are suppressed to insignificant levels. In the
(s)neutrino sector, the TeV scale effective theory is just the MSSM with 
bilinear $R$-parity violation \cite{bilinear} contained in the superpotential
terms, 
\begin{equation}
W_{\rm eff} \ni \epsilon_i\epsilon_{ab}\hat{L}_i^a\hat{H}_u^b,
\label{eq:effbi}
\end{equation}
with $\epsilon_i = h_i\frac{\langle X_2\rangle^3}{M_P^2}\sim 10^{-4}$
GeV, together with the concomitant SSB cousin of the $b_\mu$ term, 
\begin{equation}
V_{\rm soft} = \epsilon_{ab}b_{\epsilon i}\tilde{L}_i^aH_u^b + {\rm c.c.}\;,
\label{eq:B}
\end{equation}
where $\tilde{L}_i$ denotes the slepton doublet. 

The phenomenology of models with bilinear $R$-parity violation,
especially as it impacts on the neutrino sector, has been extensively
studied in the literature. Here, we will only summarize the salient
features. The sneutrino fields acquire $vevs$,
$\langle\tilde{\nu}_i\rangle
\sim\epsilon_i \sim 10^{-4}$~GeV, which lead to a mixing between the
neutral gauginos and the neutrinos. In other words, we now have a
$7\times 7$ neutral gaugino-higgsino-neutrino mass matrix that must be
diagonalized to obtain the neutrino and neutralino mass eigenstates \cite{bilinear}. 

At tree level, one linear combination of neutrinos, $\nu_3$, 
obtains a mass whose
scale is given by, 
\begin{equation}
m^{\rm tree}_\nu\sim \frac{g^2}{4}\frac{\epsilon^2}{m_{\rm SUSY}}
\label{eq:nuscale}
\end{equation}
which, for $m_{\rm SUSY}\sim 100$~GeV and $\epsilon\sim 10^{-4}$~GeV
yields a neutrino mass scale $\sim 0.1$~eV, which is of the right
magnitude to accommodate atmospheric neutrino data \cite{osci} which can
most simply be explained as $\nu_\mu-\nu_{\tau}$ oscillations with a
mass difference $\Delta m^2_{\rm atmos}=2.5\times 10^{-3}$~eV$^2$ and a
large mixing angle. A neutrino mass of $\sim 0.1$~eV is also compatible
with constraints from large scale structure formation that suggest
$m_{\nu} \alt 0.3$~eV \cite{lss}.  We must keep in mind that there is
considerable numerical lee-way in (\ref{eq:nuscale}), in that in writing
this, we have treated $\langle h_d^0\rangle$, $M_{1,2}$ and $|\mu|$ all
to be comparable and taken them to be all equal to $m_{\rm SUSY}$.

The remaining neutrinos $\nu_{1,2}$ obtain masses at the loop level,
dominantly via the diagrams shown in Fig.~\ref{fig:nuloop}, where the
bottom quark or the gaugino mass breaks the chiral symmetry. Because the
(tree-level) eigenvector for $\nu_{1,2}$ only has components in the
$\nu_e, \nu_\mu, \nu_{\tau}$ and $\th_d$ directions \cite{valle}, these
neutrinos only obtain their mass from the bottom quark-squark and
tau-stau loops in Fig.~\ref{fig:nuloop}{\it a}, whereas both bottom and
top quark-squark loops, as well as the the tau-stau loops, contribute to
the correction to $m_{\nu_3}$.
\begin{figure}
\begin{center}
\includegraphics[ width=6in]{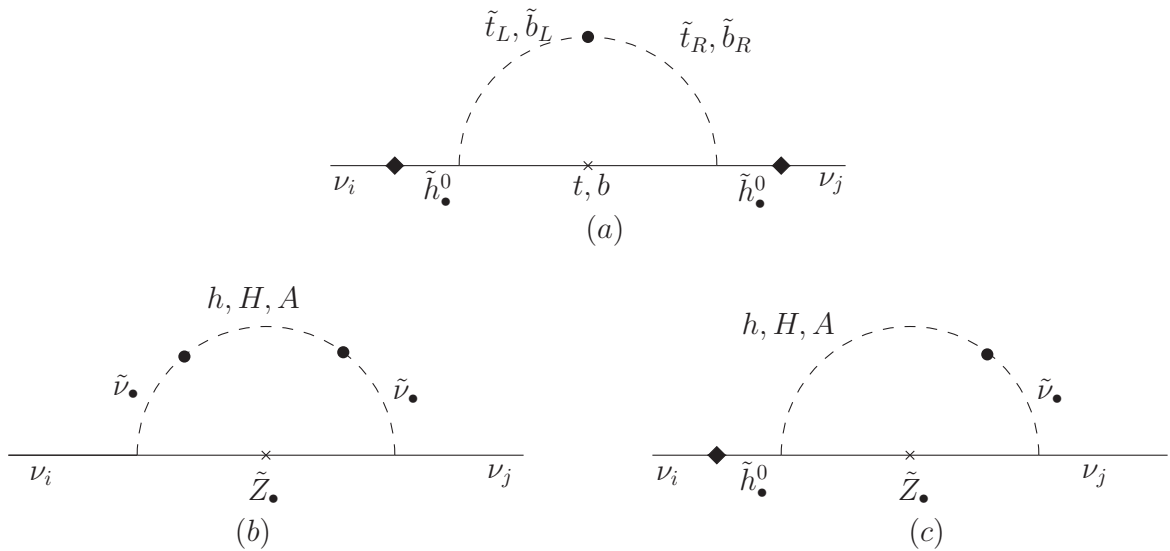}
\end{center}
\caption{One-loop diagrams that contribute to the radiative masses for
  neutrinos. There is also an analogous diagram to ({\it a}) with a
  tau-stau loop in place of the quark-squark loop. As discussed in the
  text, the diagram with the top quark-squark loop in ({\it a}) only
  contributes to the mass of $\nu_3$ (the one neutrino massive at
  tree-level) but not to the masses of $\nu_1$ and $\nu_2$.  The crosses
  denote fermion mass insertions and the solid circle denotes bilinear
  mixing between different scalars. The external neutrino lines with a
  diamond (and $\th_u/\th_d$ at the other end) symbolize that the
  neutrinos $\nu_i/\nu_j$ ($i,j=1,2,3$) couple to the quark-squark
  system via their $\th_u/\th_d$ content, and so the ``internal higgsino
  line'' is not a propagator in a Feynman diagram.}
\label{fig:nuloop}
\end{figure}
The order of magnitude of the 
contribution from Fig.~\ref{fig:nuloop}{\it a} is given by,
\begin{equation}
m_\nu^{\rm rad}\sim\frac{3}{16\pi^2}Y_q^2m_q\sin 2\theta_q\ln\left
(\frac{m_{\tq_2}^2}{m_{\tq_1}^2}\right)
\frac{\epsilon^2}{m_{SUSY}^2},
\label{eq:mnutop}
\end{equation}
where $q=t,b$, $Y_q$ is the quark Yukawa coupling, $\theta_q$ is the
corresponding squark mixing angle, and $m_{\tq_{1,2}}$, the masses of
the squarks. Detailed analyses \cite{bilinear,valle} of bilinear
$R$-parity violation have shown that the radiatively generated neutrino
mass is compatible with the solar neutrino mass scale $\Delta m^2_{\rm
solar} = 8\times 10^{-5}$~eV$^2$, obtained \cite{kamland} by
interpreting the deficit of solar neutrinos as oscillations between
neutrinos with this smaller mass difference.

We now turn to contributions from the diagrams in
Fig. \ref{fig:nuloop}{\it b} and Fig. \ref{fig:nuloop}{\it c} that have
been examined in detail in Ref.\cite{Dav-Los,Gros-Rak}, where it has
been shown that the contributions from the loops in diagrams {\it b} are
generally larger than those from diagrams {\it c}.  More importantly, a
naive estimate of these contributions obtained using loop factors,
coupling constant and mass insertions as we did for diagrams {\it a}
gives a completely wrong answer because of large cancellations between
contributions between the loop with $h, H$ and with $A$ that cause the
total contribution to vanish, both in the (unphysical) limit where
$m_h=m_A=m_H$, and in the decoupling limit where $m_A, m_H \to \infty$,
with $m_h$ and -ino masses fixed. The numerical analysis in
Ref.\cite{Gros-Rak} shows that this contribution, which scales as
$1/\cos^2\beta$, is suppressed from its naive value by a factor of about
$10^{-3}-10^{-2}$.  For $\tan\beta$ in its intermediate range $\sim 30$
yields $m_\nu^{\rm rad}\sim 10^{-2}$~eV, in general agreement with the
solar neutrino mass scale, for $b_\epsilon \sim (1-3) m_{\rm
SUSY}\epsilon$.

We have thus seen that our model naturally yields the right scale of
neutrino masses. A detailed analysis of neutrino masses and mixings, and
comparison with the observed data is beyond the scope of this paper.
Neutrino phenomenology for models with bilinear $R$-parity violation has
been examined in detail in Ref.\cite{valle} where it has been shown that
it is possible to accommodate the pattern of neutrino masses together
with the large mixing angles that provide a good fit to the solar and
the atmospheric neutrino data. 
For the ``minimal" boundary conditions used in this analysis, the
potentially large contributions from diagrams in Fig. 2{\it b} and
Fig. 2{\it c} appear to be subdominant. The analyses of Ref.\cite{adl}
illustrate that these contributions can, however, be dominant, and
further that both a normal as well as an inverted hierarchy may be
possible within this framework.
We will refer the interested reader to these studies for
  details.\footnote{It would be interesting to perform a similarly detailed
  analysis with non-degenerate sneutrinos to examine whether the model
  can also accommodate a phenomenologically viable solution with nearly
  degenerate neutrinos.}  Before moving on, we remind the reader that
  our model differs sharply from the $U(1)'$ model of Ref.~\cite{LMW} in
  that we do not include independent right-handed neutrino fields.

\subsection{New Particles at the TeV scale: Masses and Decay patterns}

We now focus our attention on the effective theory, valid at the TeV
scale, that is relevant for phenomenological analyses of new physics
signals at high energy colliders such as the LHC, or from various direct
and indirect searches for dark matter that are very topical today. This
theory is a softly broken supersymmetric $SU(3)_C\times SU(2)_L\times
U(1)_Y$ gauge theory, with the particle
content of the MSSM augmented, as we have already seen, by additional
supermultiplets of exotic particles with properties that we discuss below.  The
underlying $U(1)'$ gauge symmetry (which is spontaneously broken at the
scale $\Lambda \sim 10^{11}$~GeV), restricts the forms of both the
superpotential as well as of the SSB terms. 

\subsubsection{MSSM superpartners} We have already seen that the
non-trivial gauge kinetic function that we have introduced in
(\ref{eq:WKf}) leads to weak scale masses for MSSM gauginos. Since there
is no reason for the coefficients $f_{(\alpha)}$ to be the same, we will
generically expect that these gaugino masses are not universal.  Of
course, embedding the model into a SUSY GUT framework may yield a
common mass for the gauginos.

Although the $U(1)'$ gauge boson, which acquires a mass by the Higgs
mechanism, and the associated gaugino-higgsino states essentially
decouple from TeV scale physics, the $U(1)'$ leaves its imprint on the
scalar spectrum via a contribution from the so-called $U(1)'$ $D$-term
contributions \cite{Dterm} scalar SSB mass parameters. Thus the scalar
mass parameters are given by,
\begin{equation}
m_\bullet^2=m_{\bullet}^2({\rm high}) + {\cal Q}_\bullet \times D,
\label{eq:uipdterm}
\end{equation}
 where $m_{\bullet}^2({\rm high})$ is the SSB scalar mass parameter (in
general, non-universal if the non-minimal K\"ahler potential
terms in (\ref{eq:WKf}) are significant) for the MSSM scalars induced by
gravitational interactions, ${\cal Q}_\bullet$ is the $U(1)'$ charge of
the corresponding field as given in (\ref{eq:rel}), and $D$ is a
parameter (positive or negative) with dimension of mass squared, and a
magnitude typically around the weak scale squared. In scenarios where
the high scale SSB parameters are, for some reason, universal (remember
that scalars
of the first two generations with the same MSSM quantum numbers must be 
approximately degenerate for phenomenological reasons), the
determination of scalar masses will provide information about the
underlying $U(1)'$ charges of MSSM fields.  We remind the reader
that $m_{H_u}^2< m_{H_d}^2$ facilitates radiative electroweak
symmetry breaking.

The heavier MSSM superpartners decay as usual, mainly via their gauge
and gaugino couplings, though effects of third generation Yukawa
couplings may also be relevant \cite{BDT}. An important difference from
$R$-parity conserving models most extensively studied in the literature
is that the would-be lightest supersymmetric particle (which, for
defniteness, we will take to be the lightest neutralino, $\tz_1$) can
decay via its neutrino component that is induced by the
lepton-number-violating superpotential term. We may estimate this
component to be ${\cal O}(\epsilon_i / \mu) \sim 10^{-7}$, which
(very roughly speaking) gives a lifetime of $\alt 10^{-12}$~s [$10^{-8}$~s],
assuming that the vector bosons in $\tz_1 \to W^\pm \ell^{\mp}$ or
$\tz_1 \to Z\nu$ are real [virtual]
\cite{neutra}. Thus, except when the neutralino is lighter than $M_W$,
we would expect it to decay within the detector with, or without, a
discernable vertex separation.

\subsubsection{Exotics}

In addition to the MSSM fields, we have seen that our model includes
several TeV scale exotics. First, we have the coloured $K_i$, $K_i^\prime$
(both scalar and fermion) at the TeV scale, as we can see from the
superpotential in (\ref{eq:Wob}), remembering that $\langle
X_1\rangle=\Lambda$. Also, as we have seen in Table~\ref{tab:sol} the
scalars $X_l$ and $X_I$ (and as can be seen from (\ref{eq:Wob}) also the
fermions) acquire weak scale masses.\footnote{Actually, the situation is
more complicated than this because of the fact that there are ten
$\hat{X}_2$ fields, of which just one combination appears in the third
last term of (\ref{eq:Wob}). The spectrum that we have discussed is for
this particular combination of $\hat{X}_{2i}$ fields. The remaining nine
combinations do not affect the minimization of the scalar potential
discussed in Sec.~\ref{sec:D}, and so are more like the $\hat{Y}_{1,2}$
fields in this respect. While the scalar components of these fields get
TeV scale masses from SUSY breaking effects, the corresponding fermions
remain essentially massless. Since these remaining $\hat{X}_{2j}$ and
the $\hat{Y}_{1,2}$ fields couple to the MSSM sector or to the
$\hat{K}_i, \hat{K}_i^\prime$ fields only via the $U(1)'$ gauge interaction
(or even more weakly, via gravity), these appear to be irrelevant for
our analysis. } These particles couple to SM particles only via the very
suppressed $U(1)'$ gauge interactions (or even more weakly, via gravity
or via the Planck scale suppressed superpotential interactions), only
the coloured $\hat{K}_i$ and $\hat{K}_i^\prime$ fields are of
phenomenological interest \cite{exotic}.

We first observe that with integer hypercharge assignment for
$\hat{K}_i$ and $\hat{K}^\prime_i$, the lightest of the colour triplet
states with each integer hypercharge, be it a boson or a fermion, will
be stable since ordinary (s)quarks have fractional charges.  For the
fractional hypercharge case that we have considered, the lightest of the
$K_1/K_1^\prime$ states is stable because the colour {\it triplet} state
has charged $+\frac{1}{3}$, while the {\it anti-triplet} has the charged
$-\frac{1}{3}$. Surprisingly, the lightest of the $K_{2,3}$ and
$K_{2,3}^\prime$ states is also stable, despite the fact that
$\hat{K}_{2,3}^\prime$ and $\hat{U}^c$ have the same $SU(3)_C\times
SU(2)_L\times U(1)_Y$ quantum numbers, and so may be expected to mix
upon $U(1)'$ breaking via a term $\hat{U}^c\hat{K}_{2,3}$ that may be
induced in the superpotential when $\hat{X}_{1,2}$ or $\hat{Y}_{1,2}$
acquire $vevs$.  One can, however, readily check from the $U(1)'$
charges in (\ref{eq:rel}) that one would require {\it fractional powers}
of $\hat{X}_{1,2}$ and/or $\hat{Y}_{1,2} \times \hat{U}^c\hat{K}_{2,3}$
in the superpotential to maintain the underlying $U(1)'$
invariance.\footnote{We have checked that such a mixing is forbidden not
only for the $U(1)'$ charges in (\ref{eq:rel}) that are special to our
choice, $\frac{3{\cal Y}_1}{2{\cal X}_1}= -22$, but also for the
corresponding charges for other choices, $-1, -3, -4, -6, -10$ that we
made for this combination.}  We thus conclude that such a mixing (that
would have led to the decay of ``the $K$ states'') is not allowed. A
similar analysis shows that $\hat{K}_{2,3}^\prime \hat{D}^c\hat{D}^c$
term (which would allow the decays $K_{2,3}^\prime\to dd$, or
$\tilde{K}_{2,3}\to d\tilde{d}_R^{(*)}$ or the conjugate modes) or the 
$\hat{H}_u\hat{Q}\hat{K}_{2,3}^\prime$ terms are also
not allowed for the same set of values of $\frac{3{\cal Y}_1}{2{\cal
X}_1}$ that we have examined.

Finally, we turn to the case where $y(\hat{K}_i)=
(-\frac{2}{3},\frac{4}{3},\frac{4}{3})$ that we considered at the end of
Sec.~\ref{subsec:anomalies} so that the coloured exotics
have the same MSSM charges as the singlet quark superfields. We have
checked that even in this case mixing between $K_1$ scalars/fermions
with singlet down squarks/quarks is forbidden because the $U(1)'$ charge
of $\hat{K}_1\hat{D}^c$ equals $\frac{277}{27}{\cal X}_1$. Mixing between
$\hat{K}_{2,3}$ and $\hat{U}$ remains forbidden exactly as before since
the corresponding $U(1)'$ charges are unaffected by the flip of the weak
hypercharge of $\hat{K}_1$. We have also checked that $U(1)'$ breaking
does not induce $\hat{H}_d\hat{Q}\hat{K}_1^\prime$,
$\hat{Q}\hat{Q}\hat{K}_1$, $\hat{U}^c\hat{D}^c\hat{K}_1^\prime$,
$\hat{U}^c\hat{E}^c\hat{K}_1$ and $\hat{L}\hat{Q}\hat{K}_1^\prime$
couplings since $U(1)^\prime$ invariance can only be maintained if these
operators are multiplied by  {\it fractional powers} of  
$\hat{X}_{1,2}$ and/or $\hat{Y}_{1,2}$ fields. 
This situation thus seems to be
different from that in the models in Ref.~\cite{ma2}, \cite{LMW} and \cite{ma} 
where couplings of
the exotics to ordinary particles are possible when the exotics have the
same weak hypercharges as the singlet quarks.

\subsection{Collider Signals} Since the $R$-parity violating couplings
of MSSM superpartners are constrained by the observed neutrino masses to
be rather small, these would dominantly be pair-produced at colliders
via their gauge interactions, with cross sections as in the well-studied
$R$-parity conserving models.  They would then cascade decay
\cite{casca} to lighter sparticles as usual. The impact of the
induced $R$-parity violating couplings is that the $\tz_1$ (which we
have assumed to be the lightest MSSM superpartner) produced at the end
of the cascade is itself unstable and decays via $\tz_1\to \ell W^{(*)}$
and $\tz_1\to \nu Z^{(*)}$ as discussed above. The (real or virtual)
vector bosons decay to quarks and leptons with branching fractions given
by the SM.  In addition, the neutralino may decay via $\tz_1 \to h \
({\rm or} \ A,H) +\nu$, where $h$ dominantly decays via $b\bar{b}$.  We
refer the reader to Ref.\cite{neutra,neutbf}, where the branching fractions for
the decay of the neutralino have been examined in detail. We only
mention that while the $\eslt$ signatures may be degraded in this
$R$-parity violating scenario, the presence of charged leptons,
$b$-quarks and potential vertex gaps provide additional handles for SUSY
searches at colliders \cite{bkt,neutbf,neutra}.

The stable coloured exotics that are necessary to cancel the
$[SU(3)_C]^2\times U(1)'$ anomaly will provide the smoking-gun signature
of our scenario if they are accessible at the LHC, or at a future Very Large
Hadron Collider.  Once produced, the lighter of the colour
triplet/anti-triplet $\hat{K}_i/\hat{K}_i^\prime$ scalar or fermion
would pair up with an ordinary anti-quark/quark to form a heavy hadron,
which then decays to the {\it stable} ground state of the exotic
(s)quark--light antiquark system (or its conjugate).
For the case of integrally charged $K$'s, as well as for the
case where the colour-triplet $\hat{K}_1$ has the ``wrong'' sign of the
weak hypercharge, this stable hadron will be fractionally charged, while
in all other cases it will either be neutral or integrally charged. The
penetrating track of a slow-moving, charged heavy particle provides a
characteristic signature for the heavy charged hadron. Indeed, even in
the case that the ground state is electrically neutral, charge exchange
interactions of this hadron with the nucleon in a detector may transmute
the neutral hadron to its charged isospin partner, resulting a sudden
appearance of a track in the detector \cite{drees}. Signals from stable
quarks, squarks and gluino-hadrons in collider experiments have been
examined in the literature \cite{drees,stable}. Experiments at the
Fermilab Tevatron has carried out a search for penetrating tracks of
slow-moving heavy particles and the non-observation of any signal has led
to upper limits on the corresponding cross sections. These limits can
then be translated to lower bounds on masses of various quasi-stable
exotic particles: about 250~GeV for stable top-squarks and about
170/206~GeV for charged winos/higgsinos \cite{stable_tev}. Even with a
modest integrated luminosity of understood data, the claimed LHC reach
for gluino-hadrons/top squarks exceeds 1600/800~GeV \cite{stable_lhc}.

Before closing this section, we remark that, because the exotic
particles have negligible couplings to SM particles, the low energy
constraints on supersymmetry, {\it e.g.} from the branching ratios $b\to
s\gamma$, $b\to s\ell\bar{\ell}$, $g_\mu -2$ {\it etc.} will
be essentially as in the MSSM with the corresponding parameters.

\subsection{Cosmology and dark matter} In our 
scenario, we lose the neutralino as a thermal dark matter candidate
since it decays via $R$-parity violating couplings that give rise to
neutrino masses. While this may be viewed as a negative, it does not
exclude the model, since dark matter may reside in different sector of
the theory. Observation of a dark matter signal in direct searches for
neutralinos would rule out our model.\footnote{More precisely, it would
be an unbelievable coincidence that there is a weakly interacting
massive particle (WIMP) component of DM that has no connection with
electroweak symmetry breaking.} In this connection, we mention that it
may be possible to modify the model to allow more than one pair of
$SU(2)_L$ doublets \cite{ma} with different $U(1)^\prime$ quantum
numbers (which do not acquire a $vev$) such that a discrete subgroup
under which the new doublet transforms non-trivially is left unbroken.

A potentially more serious problem is the presence of coloured TeV scale
exotics in the $\hat{K}-\hat{K}^\prime$ sector. These would bind with
ordinary nuclei to form exotic isotopes whose expected abundance (from
thermal production in the Big Bang)\cite{wolfram} exceeds by orders of
magnitude the upper limits\cite{hemmick} on the exotic isotope fraction
for masses up to $\sim 10$~TeV: see Ref.~\cite{glashow}. These bounds may
be evaded if the reheating temperature after inflation is smaller than
the mass of the stable $K$-particles. While this is not currently
fashionable, we are not aware of any considerations that exclude this
possibility. Since the renormalizable baryon number violating operators
have extremely small couplings in our framework, low scale baryogenesis
mechanisms of Ref. \cite{baryogenesis} do not apply , and we need to
examine whether electroweak baryogenesis \cite{ewbar} can be
accommodated within the model.

\section{Summary} \label{sec:summ}
We have constructed a theoretically consistent and 
phenomenologically viable supergravity model where we impose only local
symmetries to restrict the dynamics. We fix the SUSY breaking scale by
hand to be $\Lambda \sim 10^{11}$~GeV, but otherwise assume that all
new, {\it i.e.} non-SM,  parameters
 are given by the natural values allowed by the underlying
symmetries. Our gauge group is $SU(3)_C\times SU(2)_L\times U(1)_Y\times
U(1)'$, where the $U(1)'$ gauge symmetry (which is spontaneously broken
at the intermediate scale) plays multiple roles: it serves to solve the
$\mu$ and $b_\mu$ problems, restricts the form of $R$-parity violating
interactions so that dimension three and dimension four baryon number
violating superpotential couplings, and in a class of models {\it all} 
$\Delta B=1$ couplings, are absent (solving the problem of
proton lifetime in SUSY models), and determines the order of magnitude
of the corresponding renormalizable lepton number violating
interactions. Specifically, (in the superfield basis where the fermionic
components are the $e,\mu$ and $\tau$ leptons along with the $h_d$
higgsino) trilinear lepton number violating couplings in the
superpotential are negligible, so that ``bilinear $R$-parity violation''
dominates. This, in turn, allows us to obtain a novel connection between
the scale of SUSY breaking and the mass scale of active neutrinos, that
allows us to accommodate the observed pattern of neutrino mases and
mixing angles. 

The low energy theory at the (multi)-TeV scale is the MSSM augmented by
several new fields. SUSY phenomenology would be largely that for models
with bilinear $R$-parity violation, and has been examined in the
literature. We must, however, keep in mind that often assumed
scalar universality of the mSUGRA model would generically not apply here, and
perhaps, even gaugino mass parameters may be non-universal. The only
relevant effect of $R$-parity violation in collider experiments would be
that the would-be LSP is unstable; its lepton daughters, and possible
displaced vertex would provide additional handles to enhance the
SUSY signal over SM background. Very low energy phenomenology (rare
decays, $g_\mu-2$, $CP$ violation of SM particles) is
unaltered from the MSSM because the new fields are 
extremely weakly coupled to SM particles. 

There must, however, be new colour triplet, weak iso-singlet superfields
with either the usual quark quantum numbers or exotic quantum numbers
(integrally charged quarks, or charged +$\frac{1}{3}$ quarks) which
would be copiously produced at a hadron collider if they are
kinematically accessible. Quite possibly though these fields may be at
the multi-TeV scale, and so would require a Very Large Hadron Collider
for their experimental scrutiny. The unambiguious prediction of our
model (which serves to distinguish it from other models with $U(1)$
gauge extensions) is that there are several {\it stable} coloured
particles (whether these are fermions or scalars depends on details of
parameters) which would combine with light quarks/antiquarks to form
stable hadrons.  Such hadrons would be readily discoverable at a high
energy hadron collider, where it would even be possible to determine
their mass.

Although these multi-TeV stable coloured particles provide the most striking
phenomenological signature of our model, they also cause its demise if
they are produced in the Big Bang, since their existence is excluded by
very stringent upper limits on the abundance of heavy isotopes of
hydrogen and other elements. We must, therefore assume that the
reheating temperature after inflation was low enough not to produce
these particles, and further, that the observed baryon asymmetry arises
from sphaleron effects. Finally, we do not have a WIMP candidate for DM,
so that detection of DM via direct searches would be a decisive blow to
our model. 

\section*{Acknowledgements} We thank Jason Kumar for helpful
discussions, Christoph Luhn for clarifying communications about proton
stability, and Martin Hirsch for patiently explaining the work of the
Valencia group on neutrino masses in SUSY models. This research was
supported in part by a grant from the United States Department of
Energy.


\begin{thebibliography}{99}
\bibitem{flav} Models without flavour problems generally fall into
  three categories: models where scalars with the same gauge quantum
  numbers are roughly degenerate, models where the matrices for fermions
  and corresponding sfermions are aligned, and models where scalars are
  very heavy.
\end{thebibliography}
\end{document}